\documentstyle[aps,epsf,float,preprint]{revtex}

\renewcommand{\slash}[1]{#1 \hspace{-0.55em} / }

\def\bq{\begin{eqnarray}}
\def\eq{\end{eqnarray}}
\def\be{\begin{eqnarray}}
\def\ee{\end{eqnarray}}
\def\ben{\begin{enumerate}}\def\een{\end{enumerate}}

\def\np{Nucl. Phys. }

\def\roughly#1{\mathrel{\raise.3ex\hbox{$#1$\kern-.75em
\lower1ex\hbox{$\sim$}}}}

\begin{document}

\preprint{\hfill\parbox[b]{0.3\hsize}
{ IFIC-04-54 }}

\def\bra{\langle }
\def\ket{\rangle }

\title{
Helicity-dependent
generalized parton distributions
and composite constituent quarks}

\author
{
Sergio Scopetta$^{(a,b)}$ and Vicente Vento$^{(b)}$}

\address
{\it (a)
Dipartimento di Fisica, Universit\`a degli Studi
di Perugia, via A. Pascoli
06100 Perugia, Italy, and INFN, sezione di Perugia
\\
(b)
Departament de Fisica Te\`orica,
Universitat de Val\`encia, 46100 Burjassot (Val\`encia), Spain
\\
and Institut de F\'{\i}sica Corpuscular,
Consejo Superior de Investigaciones Cient\'{\i}ficas }

\maketitle

\begin{abstract}
An approach, recently proposed to calculate the nucleon
generalized parton distributions (GPDs) in a constituent quark
model (CQM) scenario, in which the constituent quarks are taken as
complex systems, is used to obtain helicity-dependent GPDs. They
are obtained from the wave functions of the non relativistic CQM
of Isgur and Karl, {\it convoluted} with the helicity-dependent
GPDs of the constituent quarks themselves. The latter are modelled
by using the polarized structure functions of the constituent
quark, the double distribution representation of GPDs, and a
phenomenological constituent quark form factor. The present
approach permits to access a kinematical range corresponding to
both the Dokshitzer-Gribov-Lipatov-Altarelli-Parisi and the
Efremov-Radyushkin-Brodsky-Lepage regions, for small values of the
momentum transfer and of the skewedness parameter. In this
kinematical region, the present calculation represents a
prerequisite for the evaluation of cross sections relevant to
deeply virtual Compton scattering. In particular, we have
calculated the leading twist helicity-dependent GPD $\tilde H$
and, from our expressions, its general relations with the non
relativistic definition of the axial form factor and with the
leading twist polarized quark density are consistently recovered.

\end{abstract}

\section{Introduction}

Generalized parton distributions (GPDs) \cite{first,due,tre}
parametrize the non-perturbative hadron structure in hard
exclusive processes (for a recent review, see, e.g.,  \cite{md}).
Their measurement would represent a unique way to access several
crucial features of the structure of the nucleon, such as the
Angular Momentum Sum Rule of the proton \cite{tre,jaffe}, as well
as information on the structure of the proton in position space
\cite{burk}. Therefore, relevant experimental efforts to measure
GPDs, by means of exclusive electron Deep Inelastic Scattering
(DIS) off the proton, are likely to take place in the next few
years \cite{ceb}. One of the most promising processes is
represented by deeply virtual Compton scattering (DVCS)
\cite{due,tre,dvcs}.

At twist-two, four helicity-even GPDs occur.
Two of them are usually called unpolarized GPDs
and labeled $H$ and $E$. The first of them
gives, in DIS kinematics, the usual quark density.
The latter turns out to be
a spin-flip distribution, in the sense that it implies
a change of the spin of the target.
The other two are usually called polarized or
helicity-dependent GPDs. They are labeled $\tilde H$
and $\tilde E$. The first of them yields in DIS
kinematics the polarized quark density, the
second is again a spin-flip distribution.
The four of them enter the cross sections for the
DVCS process, although the two non-spin-flip
ones, $H$ and $\tilde H$, give the dominant contribution.
Reasonable estimates are necessary for the planning
of experiments.

For these quantities, several theoretical predictions have been
already produced by using different descriptions of hadron
structure: bag models \cite{meln}, soliton models \cite{pog,goe1},
light-front \cite{mill,muk} and Bethe Salpeter approaches
\cite{lukas}, and phenomenological estimates based on
parametrizations of parton distribution functions (PDFs)
\cite{freund,rad1}. Besides, an impressive effort has been devoted
to study the perturbative QCD evolution \cite{scha2} of GPDs, and
the GPDs at twist three accuracy \cite{scha1}. Chiral-odd GPDs
deserve a special mention: while they have been shown to be in
principle experimentally accessible \cite{pire}, no theoretical
estimates are available for them.

We are interested in constituent quark model (CQM) calculations.
In this framework, in order to compare model predictions with data
taken in DIS experiments, one has to evolve, according to
perturbative QCD, the leading twist component of the physical
structure functions obtained at the low momentum scale associated
with the model, from this so called ``hadronic scale'',
${\mu}_0^2$ \cite{pape,grv}, to the momentum scale of the data.
Such a procedure has proven successful in describing the gross
features of the standard PDFs by using different CQMs (see, e.g.,
\cite{trvv}). Similar expectations motivated the study of GPDs in
Ref. \cite{epj}, where a simple formalism has been proposed to
calculate the quark contribution to the GPDs from any non
relativistic model. As an illustration, results for the GPD $H_q$,
obtained in the non relativistic (NR) model of Isgur and Karl (IK)
\cite{ik} have been evolved from $\mu_0^2$ up to DIS scales, to
next to leading order (NLO) accuracy. In Ref. \cite{bpt} the same
quark contribution has been evaluated, at $\mu_0^2$,  using the
overlap representation of the GPDs \cite{kroll} in light-front
dynamics, along the lines developed in \cite{pietro}. The same
approach has been also applied to calculate the valence quark
contribution to the helicity-dependent GPDs \cite{bptv}. In all
the papers dealing with CQM listed so far \cite{epj,bpt,bptv},
only the Dokshitzer-Gribov-Lipatov-Altarelli-Parisi (DGLAP) region
of GPDs can be evaluated. To our knowledge, no CQM estimate of the
helicity-dependent GPDs has been performed in the
Efremov-Radyushkin-Brodsky-Lepage (ERBL) region. This is precisely
the argument of the present paper, where an approach recently
proposed for the unpolarized GPDs in a CQM scenario \cite{prd},
providing us with predictions also in the ERBL region, will be
extended to the helicity-dependent GPDs. In Ref. \cite{prd}, the
procedure of Ref. \cite{epj} has been extended and generalized
since, as a matter of fact, the latter, when applied to the
standard forward case, has been proven to reproduce only the gross
features of PDFs \cite{trvv}. In order to achieve a better
agreement with data, the approach has to be improved. In a series
of papers, it has been shown that unpolarized \cite{scopetta1} and
polarized \cite{scopetta2} DIS data are consistent with a low
energy scenario, dominated by complex constituent quarks inside
the nucleon, defined through a scheme suggested by Altarelli,
Cabibbo, Maiani and Petronzio (ACMP) \cite{acmp}, updated with
modern phenomenological information. The same idea has been
applied to demonstrate the evidence of complex objects inside the
nucleon \cite{psr}, analyzing intermediate energy data of electron
scattering off the proton. A similar scenario, as old as the CQM
itself \cite{mor}, has been extensively used by other groups,
starting form the concept of ``valon'' \cite{hwa}.

In Ref. \cite{prd}, the description of the unpolarized forward
case of Ref. \cite{scopetta1} has been generalized and applied to
the calculation scheme of Ref. \cite{epj}. As a result, more
realistic predictions for the GPDs have been obtained and, at the
same time, the ERBL region, not accessible before, has been
explored. In particular, the evaluation of the sea quark
contribution has become possible, so that GPDs can be calculated,
in principle, in their full range of definition. Such an
achievement will permit to estimate cross-sections that are
relevant for actual GPDs measurements, providing us with an
important tool for planning future experiments.

In here, the discussion is extended to helicity-dependent GPDs.
The description of the polarized forward case of Ref.
\cite{scopetta2} is generalized to calculate helicity-dependent
GPDs, following the path that lead to the results of Ref.
\cite{prd} starting from the calculation of the unpolarized
forward structure function in Ref. \cite{scopetta1}. As in Ref.
\cite{prd}, the proposed approach will be applied here in a NR
framework, which allows one to evaluate the GPDs only for small
values of the 4-momentum transfer, $\Delta^2$ (corresponding to
$\vec \Delta ^2 \ll m^2$, where $m$ is the constituent quark mass)
and small values also for the skewedness parameter, $\xi$. The
full kinematical range of definition of GPDs will be studied in a
followup, introducing relativity in the scheme.

The paper is structured as follows. In the second section, after
the definition of the main quantities of interest, a formula for
the helicity-dependent current quark GPD $\tilde H_q$, in a CQM
with point-like constituents, will be derived. For the same
quantity, in the third section, in a scenario dominated by
composite constituent quarks, a {\it convolution} formula in terms
of the corresponding quantity of the constituent quark and of a
polarized off-diagonal momentum distribution, will be derived in
impulse approximation (IA). Then, the helicity-dependent
constituent quark GPDs are built in the fourth section, according
to ACMP philosophy and using the Double Distribution (DD's)
representation \cite{rad1,radd,rag} of the GPDs. In the fifth
section,  results obtained by using CQM wave functions of the IK
model and our helicity-dependent constituent quark GPDs will be
shown. Conclusions will be drawn in the last section.

\section{Calculations of helicity-dependent GPDs
in models with point-like quarks}

Let us consider a hard exclusive process, where the absorption of
a high-energy virtual photon by a quark in a hadron target is
followed by the emission of a particle to be later detected;
finally, the interacting quark is reabsorbed back into the
recoiling hadron. If the emitted and detected particle is, for
example, a real photon, the so called Deeply Virtual Compton
Scattering process takes place  \cite{due,tre,pog}. We adopt here
the formalism used in Ref. \cite{jig}. Let us think of a nucleon
target, with initial (final) momentum and helicity $P(P')$ and
$s(s')$, respectively. The GPDs $\tilde H_q(x,\xi,\Delta^2)$ and
$\tilde E_q(x,\xi,\Delta^2)$ are defined through the expression
\begin{eqnarray}
\label{eq1}
\tilde F^q_{s's}(x,\xi,\Delta^2) & = &
{1 \over 2} \int {d \lambda \over 2 \pi} e^{i \lambda x}
\bra P' s' | \, \bar \psi_q \left(- {\lambda n \over 2}\right)
\slash{n} \, \gamma_5
\, \psi_q \left({\lambda n \over 2} \right) | P s \ket  =
\nonumber
\\
& = & \tilde H_q(x,\xi,\Delta^2) {1 \over 2 }\bar U(P',s')
\slash{n} \, \gamma_5 U(P,s) \, +
\tilde E_q(x,\xi,\Delta^2) {1 \over 2} \bar U(P',s')
{\Delta \cdot n \, \gamma_5 \, \over 2M} U(P,s)~,
\end{eqnarray}
\\
where
$\Delta=P^\prime -P$
is the 4-momentum transfer to the nucleon,
$\psi_q$ is the quark field and M is the nucleon mass.
It is convenient to work in
a system of coordinates where
the photon 4-momentum, $q^\mu=(q_0,\vec q)$, and $\bar P=(P+P')/2$
are collinear along $z$.
The $\xi$ variable in the arguments of the GPDs
is the so called ``skewedness'', parametrizing
the asymmetry of the process. It is defined
by the relation $\xi = - n \cdot \Delta/2$,
where $n$
is a light-like 4-vector
satisfying the condition $n \cdot \bar P = 1$.
As explained in \cite{tre}, GPDs describe the amplitude for
finding a quark with momentum fraction $~~x+\xi$ (in the Infinite
Momentum Frame) in a nucleon with momentum $(1+\xi) \bar P$ and
replacing it back into the nucleon with a momentum transfer
$\Delta$. Besides, when the quark longitudinal momentum fraction
$x$ of the average nucleon momentum $\bar P$ is less than $-\xi$,
GPDs describe antiquarks; when it is larger than $\xi$, they
describe quarks; when it is between $-\xi$ and $\xi$, they
describe $q \bar q$ pairs. The first and second $x$ intervals  are
commonly referred to as the DGLAP region and the third as the ERBL
region \cite{jig}, following the pattern of evolution in the
factorization scale. One should keep in mind that, besides the
variables $x,\xi$ and $\Delta^2$ explicitly shown, GPDs depend, as
the standard PDFs, on the momentum scale $Q^2$ at which they are
measured or calculated. To make the presentation easier looking,
we omit the latter dependence unless when specifically needed. The
values of $\xi$ which are possible for a given value of $\Delta^2$
are: \bq 0 \le \xi \le \sqrt{- \Delta^2}/\sqrt{4 M^2-\Delta^2}~.
\label{xim} \eq The known constraints of $\tilde
H_q(x,\xi,\Delta^2)$ are:

i) the so called ``forward'' limit, $P^\prime=P$, i.e.,
$\Delta^2=\xi=0$, yields the usual polarized PDFs \bq \tilde
H_q(x,0,0)=\Delta q(x)~; \label{i)} \eq

ii) the integration over $x$, yields the contribution of the quark
of flavor $q$ to the axial form factor (ff) of the target $A$,
called hereafter $G_A^q(\Delta^2)$: \bq \int dx \tilde
H_q(x,\xi,\Delta^2) = G_A^q(\Delta^2)~; \label{ii)} \eq

iii) the polynomiality property \cite{jig}, according to which the
$x$-integrals of $x^n \tilde H^q$ and of $x^n \tilde E^q$ are
polynomials in $\xi$ of order $n+1$.

In \cite{epj,prd}, the IA expression for the unpolarized
$H_q(x,\xi,\Delta^2)$, suitable to perform CQM calculations, was
obtained. In a similar fashion, we derive next a formula for the
helicity-dependent GPD $\tilde H_q(x,\xi,\Delta^2)$ from the
definition (\ref{eq1}).

Using the light-cone spinor definitions as given in the appendix B
of \cite{md}, and defining: $k^+=(k_0+k_3)/\sqrt{2}, \vec k_\perp
= (k_1,k_2)$, for the light-cone helicity combination $s's={1
\over 2} {1 \over 2} =++$ one obtains
\begin{eqnarray}
\tilde F^q_{++}(x,\xi,\Delta^2) = {\sqrt{1 - \xi^2} }
\tilde H_q(x,\xi,\Delta^2) - {\xi^2 \over
\sqrt{1 - \xi^2}} \tilde E_q(x,\xi,\Delta^2)~,
\end{eqnarray}
so that, for $\xi^2 \ll 1$:

\begin{eqnarray}
\tilde F^q_{++}(x,\xi,\Delta^2)  =
{\tilde H_q(x,\xi,\Delta^2)}
- \xi^2
\left( {1 \over 2} \tilde H_q(x,\xi,\Delta^2)
+ \tilde E_q(x,\xi,\Delta^2)  \right)
+ O(\xi^4)~,
\end{eqnarray}

i.e.

\begin{eqnarray}
\tilde F^q_{++} (x,\xi,\Delta^2) = \tilde H_q (x,\xi,\Delta^2)
+ O(\xi^2)~.
\label{hf++}
\end{eqnarray}

According to this equation, in order to obtain the GPD $\tilde
H_q(x,\xi,\Delta^2)$ for $\xi^2 \ll 1$ one has to evaluate $\tilde
F_{++}^q$, starting from its definition, Eq (\ref{eq1}). Using
light-cone quantized quark fields in the l.h.s., whose creation
and annihilation operators, $b^\dag(k)$ and $b(k)$, obey the
commutation relation \bq \{ b(k'),b^\dag(k)\} = (2 \pi)^3 2 k^+
\delta(k'^+ - k^+) \delta^2(\vec k_\perp - \vec k_\perp')~, \eq
and using light-cone states normalized as
\bq \bra P' | P \ket =
(2 \pi)^3 2 P^+ \delta(P'^+ - P^+) \delta^2(\vec P_\perp - \vec
P_\perp')~, \eq
one obtains, for $x > \xi$ \cite{jig}
\bq \tilde
F^q_{++} (x,\xi,\Delta^2) = {1 \over 2 \bar P^+ V} \int {d^2
k_\perp \over 2 \sqrt{ |x^2 - \xi^2|} (2 \pi)^3 } \bra b^\dag (k +
\Delta) b(k) \ket ~, \label{fqpp} \eq
where
\bq \bra b^\dag(k')
b(k) \ket= \sum_\lambda sign(\lambda) \bra P'+ |
b_\lambda^\dag((x-\xi)\bar{P}^+, {\vec k_\perp'} )
b_\lambda((x+\xi)\bar{P}^+, {\vec k_\perp})| P+ \ket~, \eq
and $V$
is a volume factor. Since we want to obtain only the quark
contribution to $\tilde H_q$, the only one which can be evaluated
in a CQM with three valence, point-like quarks, we are interested
here in the $x > \xi$ region.

Eq. (\ref{fqpp}) can be written (see \cite{prd}
where all the steps of a similar
derivation are explicitely given):
\bq
\tilde F^q_{++} (x,\xi,\Delta^2) =
{1 \over 2 \bar P^+ V} \int
{d^2 k_\perp d k^+ \over 2 \sqrt{ k^+
k'^+} (2 \pi)^3 }
\delta \left( {k^+ \over \bar P^+} - (x+\xi) \right)
\bra b^\dag (k + \Delta) b(k)  \ket~.
\label{fqpp1}
\eq

In a NR framework, states and creation and annihilation operators
have to be normalized according to
\bq \bra \vec P' | \vec P \ket
= (2 \pi)^3 \delta(P'^+ - P^+) \delta(\vec P'_\perp - \vec
P_\perp)~ \eq
and
\bq \{ b(k'^+,\vec k_\perp),b^\dag (k^+, \vec
k'_\perp)\} = (2 \pi)^3 \delta(k'^+ - k^+) \delta(\vec k'_\perp -
\vec k_\perp)~ \eq
respectively. As a consequence, one has to
consider that \cite{muld}:
\bq | P \ket \rightarrow \sqrt{2
P^+}|\vec P \ket~, \eq \bq b(k) \rightarrow \sqrt{2 k^+}b(k^+,\vec
k_\perp)~, \eq
so that in Eq. (\ref{fqpp1}), in terms of the new
states and fields, one has to perform the substitution
\bq \bra
b^\dag (k + \Delta) b(k) \ket & = & \sum_\lambda \, sign(\lambda)
\, \bra P'+ | b_\lambda^\dag((x-\xi)\bar{P}^+, \vec k_\perp + \vec
\Delta_\perp) b_\lambda((x+\xi)\bar{P}^+, \vec k_\perp )| P + \ket
\rightarrow \nonumber
\\
& \rightarrow &
2 \sqrt{1 - \xi^2}\bar P^+
\sqrt{2 k'^+ 2 k^+} \times
\nonumber
\\
& & \times
\sum_\lambda \, sign(\lambda) \,
\bra \vec P' |
b_\lambda^\dag(k^+ + \Delta^+, \vec k_\perp + \vec \Delta_\perp )
b_\lambda(k^+, \vec k_\perp )| \vec P \ket~,
\label{newnorm}
\eq
where the relation
$2 \sqrt{1 - \xi^2}\bar P^+ = \sqrt{2 P^+} \sqrt{2 P'^+}$
has been used.

Now, inserting Eq. (\ref{newnorm}) in Eq. (\ref{fqpp1}),
one gets
\bq
\tilde
F_{++}^q(x,\xi,\Delta^2)
& = &
{ 1 \over V} \int
{d^2 k_\perp d k^+ \over 2 \sqrt{ k^+
k'^+} (2 \pi)^3 }
\delta \left( {k^+ \over \bar P^+} - (x+\xi) \right)
\, \sqrt{1 - \xi ^2} \, 2 \sqrt{ k^+
k'^+}
\times
\nonumber
\\
& \times &
\sum_\lambda \, sign(\lambda) \,
\bra \vec P' |
b_\lambda^\dag(k^+ + \Delta^+, \vec k_\perp + \vec \Delta_\perp )
b_\lambda(k^+, \vec k_\perp )| \vec P \ket  =
\nonumber
\\
& = &
{1 \over V} \int
{d^2 k_\perp d k^+ \over (2 \pi)^3 }
\delta \left( {k^+ \over \bar P^+} - (x+\xi) \right)
\times
\nonumber
\\
& \times &
\sum_\lambda \, sign(\lambda) \,
\bra \vec P' |
b_\lambda^\dag(k^+ + \Delta^+, \vec k_\perp + \vec \Delta_\perp)
b_\lambda(k^+, \vec k_\perp )| \vec P \ket + O(\xi^2)~.
\label{intm}
\eq

The constituent quarks with mass $m$ are taken to be on shell,
$k_0 = \sqrt{ \vec k^2 + m^2}$, thus
\bq d^2 k_\perp d k^+
\rightarrow \left( { k^+ \over k_0} \right) d \vec k~, \eq
and
from Eq. (\ref{intm}) and Eq. (\ref{hf++}), one gets
\bq \tilde
H_q(x,\xi,\Delta^2) & = & \int {d \vec k } \,\, \delta \left( {k^+
\over \bar P^+} - (x+\xi) \right) \times \nonumber
\\
& \times &
\left [
{ 1 \over (2 \pi)^3 V}
{k^+ \over k_0}
\sum_\lambda \, sign(\lambda) \,
\bra \vec P' |
b_\lambda^\dag(k^+ + \Delta^+, \vec k_\perp + \vec \Delta_\perp)
b_\lambda(k^+, \vec k_\perp )| \vec P \ket \right] + O(\xi^2) =
\label{int_1}
\\
& = &
\int  d \vec k \,\,
\delta \left ( {k^+ \over \bar P^+} - (x+\xi) \right )
\left[\, \tilde n_q ( \vec k, \vec k + \vec \Delta) + O \left(
{\vec k ^2 \over m^2}, {(\vec k + \vec \Delta)^2 \over m^2}
\right)  \, \right] + O(\xi^2)~. \label{inter} \eq
In the last
step, the definition of a (non-diagonal) polarized NR momentum
distribution, $\,\tilde n_q ( \vec k, \vec k + \vec \Delta)$, has
been used, together with the fact that a NR momentum distribution
describes the probability of finding a constituent of momentum
$\vec k$ in a given system up to terms of order ${\vec k^2 \over
m^2}$ \cite{fs}.

So far, we have shown that, in a NR CQM, the GPD $\tilde H_q(x,\xi,\Delta^2)$
can be calculated, for $\xi^2 \ll 1$, $\vec k^2 \ll m^2$ and
$(\vec k + \vec \Delta)^2 \ll m^2$ (which means, in turn,
$\vec \Delta^2 \ll m^2$), through the following expression:
\begin{eqnarray}
\tilde H_q(x,\xi,\Delta^2) =
\int  d \vec k \,
\delta \left( {k^+ \over \bar P^+} - (x+\xi) \right)
\, \tilde n_q ( \vec k, \vec k + \vec \Delta)
+ O \left( \xi^2, {\vec k ^2 \over m^2}, {\vec \Delta^2 \over m^2}
\,\right)
~.
\label{usu}
\end{eqnarray}
The above equation, corresponding to Eq. (8) in \cite{epj} if
polarization is taken into account, permits the calculation of
$\tilde H_q(x,\xi,\Delta^2)$ in any CQM, and it naturally verifies
some of the properties of the GPDs. In fact, the polarized quark
density, $\Delta q(x)$, as obtained  by analyzing, in IA, the DIS
off the nucleon (see, e.g., \cite{rope}), assuming that the
interacting quark is on-shell, is recovered in the forward limit,
where $\Delta^2=0, \xi=0$:
\begin{eqnarray}
\Delta q(x) = \tilde H_q(x,0,0)=
\int
d \vec k
\, \tilde n_q(\vec k) \, \delta \left ( x - {k^+ \over  P^+}
\right )~,
\label{hf}
\end{eqnarray}
so that the constraint Eq. (\ref{i)}) is fulfilled. In the above
equation, $\tilde n_q(\vec k)$ is the NR polarized momentum
distribution of the quarks in the nucleon and, integrating Eq.
(\ref{usu}) over $x$, one obtains the contribution of the quark of
flavor $q$ to the IA definition of the axial ff, so that Eq.
(\ref{ii)}) is verified.

The polynomiality condition is formally fulfilled
by the GPD defined in Eq. (\ref{usu}), although the present
accuracy of the model, explicitly written in the latter equation,
does not allow to really check polynomiality, due to the
effects of the Munich Symmetry \cite{mun}, which excludes
$O(\xi)$ contributions.

With respect to Eq. (\ref{usu}),
a few caveats, already listed in
Ref. \cite{prd} for the unpolarized case, are necessary.

i) One should keep in mind that
Eq. (\ref{usu}) is a NR result, holding for $\xi^2 \ll 1$,
under the conditions $\vec k^2 \ll m^2$,
$\vec \Delta^2 \ll m^2$.
If one wants to treat more general processes,
the NR limit should be relaxed by
taking into account relativistic corrections.
In this way,
at the same time, an expression to
evaluate $\tilde E_q(x,\xi,\Delta^2)$ could be
obtained.

ii) The Constituent Quarks are point-like objects.

iii) If use is made of a CQM, containing only
constituent quarks (and also antiquarks in the case of mesons),
only the quark (and antiquark) contribution to the
GPDs can be evaluated, i.e.,
only the region $x \geq \xi$
(and also $x \leq -\xi$ for mesons) can be explored.
In order to introduce the study of the ERBL region
($ - \xi \leq x \leq \xi$), so that
observables like cross-sections, spin asymmetries and so on
can be calculated,
the model has to be improved.

iv) In actual calculations, the evaluation of
Eq. (\ref{usu}) requires the choice of a reference frame.
In the following, the Breit Frame will be chosen, where one has
$\Delta^2 = - \vec \Delta^2$ and, in the NR limit we are studying,
one finds $\sqrt{2} \bar P ^+ \rightarrow M$. It happens therefore that,
in the argument of the $\delta$ function
in Eq. (\ref{usu}),
the $x$ variable for the valence quarks
is not defined in its natural support, i.e. it can be
larger than 1 and smaller than $\xi$.
Although the
support violation is small for the calculations that will be shown here,
it has to be reported as a drawback of the approach.

The issue iii) will be discussed in the next sections,
by relaxing the condition ii) and allowing for a finite size
and composite structure of the constituent quark.

\section{Helicity- dependent GPDs in a Constituent Quark Scenario}

In the standard forward case,
the procedure described in the previous section,
has been proven to be able to
reproduce the gross features of PDFs
\cite{trvv}.
In order to achieve a better agreement with data, the approach
has to be enriched.

In previous papers, it has been shown that
unpolarized
\cite{scopetta1} and polarized \cite{scopetta2} DIS data
are consistent with a low energy scenario
dominated by
complex systems of point-like partons.
The forward parton distributions
are therefore given by a convolution between constituent
quark momentum distributions
and constituent quark structure functions.
The latter quantities were obtained by using updated phenomenological
information in a scenario firstly suggested by Altarelli,
Cabibbo, Maiani and Petronzio (ACMP) already in the seventies
\cite{acmp}.
Following the same idea, in Ref. \cite{prd}
a model for the reaction mechanism of an off-forward process,
such as DVCS, where GPDs could be measured,
has been proposed.
As a result, a convolution formula giving
the proton $H_q$ GPD in terms of a constituent quark off-forward momentum
distribution,
$H_{q_0}$,
and of a GPD of the constituent quark $q_0$ itself,
$H_{q_0q}$,
has been derived.
In here, a similar program will be developed for the
helicity-dependent GPD $\tilde H_q$.

It is assumed that the hard scattering with the virtual photon
takes place on a parton of a spin $1/2$ target, such as
the proton, made of
complex constituents.
The scenario is depicted in Fig. 1 for the
special case of DVCS, in the handbag
approximation. One parton ({\sl{current}} quark)
with momentum $k$,
belonging to a given constituent
of momentum p, interacts with the probe
and it is afterwards reabsorbed, with momentum
$k+\Delta$, by the same constituent,
without further re-scattering with the recoiling system
of momentum $P_R$.
We suggest here an analysis of the process
which is quite similar to the usual IA approach
to DIS off nuclei \cite{fs}, recently applied
also to nuclear GPDs \cite{prc}.

In the class of frames chosen in section 2,
and in addition to the kinematical variables,
$x$ and $\xi$
one needs a few more to describe the process.
In particular,
$x'$ and $\xi'$, for the ``internal''
target, i.e., the constituent quark, have to be introduced.
These quantities can be obtained defining the ``+''
components of the momentum $k$ and $k + \Delta$ of the struck parton
before and after the interaction, with respect to
$\bar P^+$ and $\bar p^+ = {1 \over 2} (p + p')^+$:
\begin{eqnarray}
k^+ & = & (x + \xi ) \bar P^+ =  (x' + \xi') \bar p^+~,
\\
(k+\Delta)^+ & = &
(x - \xi ) \bar P^+ =  (x' - \xi') \bar p^+~.
\end{eqnarray}
From the above expressions, $\xi'$ and $x'$ are immediately obtained as
\begin{eqnarray}
\xi' & = & - { \Delta^+ \over 2 \bar p^+}
\\
x' & = & {\xi' \over \xi} x
\label{x1}
\end{eqnarray}
and, since $\xi = - \Delta^+ / (2 \bar P^+)$,
if $\tilde z = p^+/P^+$, one also has
\begin{eqnarray}
\xi ' = {\xi \over \tilde z( 1 + \xi ) - \xi}~.
\label{xi1}
\end{eqnarray}

In order to derive a convolution formula in IA for $\tilde H_q$,
the procedure used for standard DIS will be adopted \cite{fs}.
First, in Eq. (\ref{int_1}),
two complete sets of states, corresponding
to the interacting particle and to the recoiling
system, are properly inserted
to the left and right-hand sides of the quark operator:
\begin{eqnarray}
{\tilde H_q(x,\xi,\Delta^2)} & = &
{\bra P'S |} \,\,\sum_{ P_R',S_R', p',s'}
{ \{ | P_R' S_R' \ket | p' s' \ket \}}
\{ \bra  P_R' S_R' |  {\bra  p' s'|} \}
\nonumber
\\
& &
\int
{d \vec k \over (2 \pi )^3 }
{ k^+ \over k_0}
\delta \left( {k^+ \over \bar P^+} - (x+\xi) \right)
{ 1 \over V}
\sum_{\lambda} \, sign(\lambda) \,
b_{q,\lambda}^\dag(k^+ + \Delta^+, \vec k_\perp + \vec \Delta_\perp)
b_{q,\lambda}(k^+, \vec k_\perp )
\nonumber
\\
& &
\sum_{ P_R,S_R,p,s}
\{ |  P_R S_R \ket {| p s \ket \}}
{\{ \bra  P_R S_R |  \bra  p s| \}
\,\,\,\,|   P S  \ket}~;
\label{iaint}
\end{eqnarray}
second,
since eventually use has to be made of NR wave functions, a
state $| \vec p \ket $
has to be normalized in a NR manner:
\bq
\bra \vec p' | \vec  p \ket = (2 \pi)^3
\delta(\vec p' - \vec p)~,
\label{norr}
\eq
so that in Eq. (\ref{iaint}) one has to perform
the substitution:
\bq
| p > \rightarrow \sqrt{{p^+ \over p^0} }| \vec p >~.
\eq
Since, using IA in the intrinsic frame of the target one has,
for the NR states:
\begin{eqnarray}
{\{ \bra  \vec P_R S_R |  \bra  \vec p s| \}
| \vec P S \ket} = {\bra \vec P_R S_R, \vec p s
| \vec P S \ket } (2 \pi)^3 \delta^3 (\vec P - \vec P_R - \vec p)
\delta_{S,S_R\,s}~,
\nonumber
\end{eqnarray}
a convolution formula is readily obtained from Eq. (\ref{iaint}):
\begin{eqnarray}
\tilde H_q(x,\xi,\Delta^2) & \simeq &
\sum_{q_0} \int dE \int d \vec p
\,
\sqrt{p^+ (p+\Delta)^+ \over p_0 (p + \Delta)_0}
\tilde P_{q_0}(\vec p, \vec p + \vec \Delta, E )
{\xi' \over \xi}
\tilde H_{q_0q}(x',\xi',\Delta^2) +
{\cal{O}}
\left ( \xi^2 \right ) =
\label{flux}
\\
& =  &
\sum_{q_0} \int dE \int d \vec p
\,
[ \tilde P_{q_0}(\vec p, \vec p + \vec \Delta, E ) +
{\cal{O}}
( {\vec p^2 / M^2},{\vec \Delta^2 / M^2}) ]
\nonumber
\\
& \times &
{\xi' \over \xi}
\tilde H_{q_0 q}(x',\xi',\Delta^2) +
{\cal{O}}
\left ( \xi^2 \right )~.
\label{spec}
\end{eqnarray}
In the above equations, the kinetic energies of the residual
system and of the recoiling target have been neglected, and
$\tilde P_{q_0} (\vec p, \vec p + \vec \Delta, E )$ is
a spin dependent, off-diagonal spectral function
for the constituent quark in the target:
\begin{eqnarray}
\tilde P_{q_0}(\vec p, \vec p + \vec \Delta, E)  & = &
{1 \over (2 \pi)^3}
\sum_{R,s} \, sign(s) \,
\bra \vec P'M | (\vec P - \vec p) S_R, (\vec p + \vec \Delta) s\ket
\bra (\vec P - \vec p) S_R,  \vec p s| \vec P M \ket
\times
\nonumber
\\
& \times &
\, \delta(E - E^*_R)~.
\label{spectral}
\end{eqnarray}

Besides, the quantity

\bq
\tilde H_{q_0q}(x',\xi',\Delta^2)
& = &
\int
{d \vec k } \,\,
\delta \left( {k^+ \over \bar p^+} - (x'+\xi') \right)
\nonumber
\\
& \times &
{ 1 \over (2 \pi)^3 V}
{k^+ \over k_0}
\sum_\lambda \, sign(\lambda) \,
\bra p' |
b_{q,\lambda}^\dag(k^+ + \Delta^+, \vec k_\perp + \vec \Delta_\perp)
b_{q,\lambda}(k^+, \vec k_\perp ) | p \ket
=
\nonumber
\\
& = &
{\xi \over \xi'}
\int
{d \vec k } \,\,
\delta \left(
{k^+ \over \bar P^+} -
(x +\xi) \right)
\nonumber
\\
& \times &
{ 1 \over (2 \pi)^3 V}
{k^+ \over k_0}
\sum_\lambda \, sign(\lambda) \,
\bra p' |
b_{q,\lambda}^\dag(k^+ + \Delta^+, \vec k_\perp + \vec \Delta_\perp)
b_{q,\lambda}(k^+, \vec k_\perp ) | p \ket
\label{gpdb}
\eq
is, according to Eq. (\ref{int_1}),
the helicity-dependent GPD of the constituent quark $q_0$
up to terms of order $O(\xi^2)$, and in the above equation
use has been made
of Eqs. (\ref{x1}) and (\ref{xi1}).

The delta function in Eq (\ref{spectral})
defines $E$, the removal energy, in terms of
$E^*_R$, the excitation energy of the recoiling system.

Concerning Eqs (\ref{spec}--\ref{gpdb}), two comments are in order.

The first concerns the accuracy of the actual calculations
which will be presented. In the following, NR wave functions
will be used to evaluate Eq. (\ref{spec}),
so that the accuracy of the calculation is
of order
${\cal{O}}
\left ( {\vec p^2 / M^2},{\vec \Delta^2 / M^2} \right )$.
In fact, if use is made of NR wave
functions to calculate the non-diagonal spectral function,
the result holds for
$\vec p^2, ( \vec p + \vec \Delta)^2 << M^2$.
The same constraint can be written
$\vec p^2, \vec \Delta^2 << M^2$.

Second, from the study of forward DIS off nuclei, it is known
that,
in going from a covariant formalism
to one where use can be made of the usual
NR wave functions, one has to
keep the correct normalization
\cite{fs}. This procedure
leads to the appearance of the ``flux factor'',
represented in Eq. (\ref{flux}) by the expression
$\sqrt{ p^+ (p + \Delta)^+ / p_0 (p + \Delta)_0 }$
(which gives, in forward DIS, the usual $p^+ / p_0$ term).
This factor gives one at order
${\cal{O}}
\left ( {\vec p^2 / M^2},{\vec \Delta^2 / M^2} \right )$,
which is the accuracy of the present analysis, and
it will not be included in the
calculation described below.
If one wants to be predictive in more general processes
at higher momentum transfer,
the accuracy of Eq. (\ref{spec}) is not good enough any more and
the calculation, in order to be consistent, has to be performed
taking into account relativistic corrections.

Eq. (\ref{spec}) can be written in the form
\begin{eqnarray}
\tilde H_{q}(x,\xi,\Delta^2) & = &
\sum_{q_0} \int d E
\int d \vec p
\, \tilde P_{q_0}(\vec p, \vec p + \vec \Delta) {\xi' \over \xi}
\tilde H_{q_0 q}(x',\xi',\Delta^2) = \nonumber
\\
& = &
\sum_{q_0} \int_x^1 {dz \over z}
\int d E
\int d \vec p
\, \tilde P_{q_0}(\vec p, \vec p + \vec \Delta)
\delta \left( z - { \xi \over \xi' }  \right)
\tilde H_{q_0 q} \left( { x \over z }, { \xi \over z },\Delta^2 \right)~.
\label{pnk}
\end{eqnarray}
Taking into account that
\begin{equation}
z - { \xi \over \xi'} = z - [ \tilde z ( 1 + \xi  ) - \xi ]
= z + \xi - { p^+ \over P^+} ( 1 + \xi )
= z + \xi  - { p^+ \over \bar P^+}~,
\end{equation}
Eq. (\ref{pnk}) can also be written in the form:
\begin{eqnarray}
\tilde H_{q}(x,\xi,\Delta^2) =
\sum_{q_0} \int_x^1 { dz \over z}
\tilde H_{q_0}(z, \xi ,\Delta^2 )
\tilde H_{q_0q} \left( {x \over z},
{\xi \over z},\Delta^2 \right)~,
\label{main}
\end{eqnarray}
where
\begin{equation}
\tilde H_{q_0}(z, \xi ,\Delta^2 ) =
\int d E
\int d \vec p
\, \tilde P_{q_0}(\vec p, \vec p + \vec \Delta)
\delta \left( z + \xi  - { p^+ \over \bar P^+ } \right)~.
\label{hq0}
\end{equation}

One should notice that
Eqs. (\ref{main}) and (\ref{hq0}) or, which is the same,
Eq. (\ref{spec}), fulfill the constraint $i)-iii)$ previously listed.

The constraint $i)$, i.e. the forward limit
of GPDs, is verified.
In fact, by taking
the forward limit ($\Delta^2 \rightarrow 0, \xi \rightarrow 0$)
of Eq. (\ref{main}), one gets the
expression which is usually found,
for the polarized parton distribution $\Delta q(x)$, in the IA analysis of
polarized DIS off polarized three-body systems
made of composite constituents
(see, i.e.,
\cite{csps} for the $^3$He target)
\begin{eqnarray}
\Delta q(x) =  \tilde H_q(x,0,0) =
\sum_{q_0} \int_x^1 { dz \over z}
g_{q_0}(z) \,
\Delta q_{q_0}\left( {x \over z}\right)~.
\label{mainf}
\end{eqnarray}
In the latter equation,
\begin{equation}
g_{q_0}(z) = \tilde H_{q_0}(z, 0 ,0) =  \int d E \int d \vec p
\, \tilde P_{q_0}(\vec p,E)
\delta\left( z - { p^+ \over  P^+ } \right)
\label{hq0f}
\end{equation}
is the light-cone momentum distribution of the constituent $q_0$
in the nucleon, $\Delta q_{q_0}(x)= \tilde H_{q_0q}( x , 0, 0)$
is the distribution
of the quark of flavour $q$
in the constituent $q_0$ and $\tilde P_{q_0}(\vec p, E)$,
the $\Delta^2 \longrightarrow 0$ limit of
Eq. (\ref{pnk}), is the
one body spin dependent spectral function, labeled
$P_{||}$ in Ref. \cite{cps,csps}.
If the system were a three-body nucleus,
the constituents would be nucleons and Eqs.
(\ref{main})-(\ref{hq0f})
would coincide with the
corresponding ones obtained in Ref. \cite{csps}
for the polarized $^3$He target.

The constraint $ii)$, i.e. the $x-$integral of the GPD
$\tilde H_q$, is also naturally
fulfilled. In fact, by $x-$integrating Eq. (\ref{main}),
one easily obtains:
\begin{eqnarray}
\int dx \tilde H_q (x,\xi,\Delta^2) & = & \sum_{q_0}
\int dx \int {dz \over z} \tilde H_{q_0}(z,\xi,\Delta^2)
\tilde H_{q_0q} \left ( { x \over z}, {\xi \over z},
\Delta^2 \right ) =
\nonumber
\\
& = &
\sum_{q_0}
\int d x' \tilde H_{q_0 q} (x',\xi',\Delta^2) \int d z
\tilde H_{q_0}(z,\xi,\Delta^2) =
\nonumber
\\
& = &
\sum_{q_0}
G^q_{q_0}(\Delta^2)
G^{q_0}_p(\Delta^2)
= G^q_p(\Delta^2)~.
\label{ffc}
\eq
In the equation above,
$G_p^q(\Delta^2)$ is the
contribution,
of the quark of flavour $q$,
to the
proton axial ff;
$G^q_{q_0}(\Delta^2)$ is the contribution,
of the quark of flavour $q$,
to the constituent quark $q_0$ axial ff;
$G_{q_0}^p(\Delta^2)$ is a
proton axial ``pointlike ff'', which
would represent the contribution of the quark $q_0$ to the
axial ff of the proton if $q_0$ were point-like.
$G_p^{q_0}(\Delta^2)$ is given, in the present approximation, by
\bq
G_p^{q_0}(\Delta^2) = \int dE \int d \vec p
\, \tilde P_{q_0}(\vec p, \vec p + \vec \Delta, E)
= \int dz \, \tilde H_{q_0}(z,\xi,\Delta^2)~.
\label{ffp}
\eq

Eventually the polynomiality, condition $iii)$,
is formally fulfilled by Eq. (\ref{spec}), although
one should always remember that it is a
result of order ${\cal{O}}(\xi^2)$,
so that high moments cannot be really checked.

Summarizing, an IA {\it convolution} formula has been derived,
valid for any helicity-dependent GPD $\tilde H_q$ of a spin $1/2$
hadron target made of three spin $1/2$ constituents, Eq.
(\ref{spec}) (or, which is equivalent, Eq. (\ref{main})), at order
${\cal{O}}\left ({\vec p^2 \over M^2}, {\vec \Delta^2 \over M^2},
\xi^2 \right )$, in terms of a spin-dependent, non-diagonal
spectral function, Eq. (\ref{spectral}), and of the
helicity-dependent GPD $\tilde H_{q_0q}$ of the internal target.
Convolution formulae are usually obtained in describing
quantities, such as structure functions, which have a
probabilistic interpretation. This is the case of Eq.
(\ref{mainf}). However, GPDs represent amplitudes and not
probability densities. Our convolution equation for GPDs,
obtained as the result of a mathematical procedure, also used by
several authors in the context of the IA in nuclear GPDs
\cite{cano,str,prc}, has a different meaning. It involves wave
functions evaluated for different values of the momentum of one of
the particles (one body off diagonal matrix elements). Besides, in
the derivation, the simple structure of the final result, Eq.
(\ref{main}), is due to the restrictions imposed in the kinematics
($\Delta^2 \ll m^2, \xi^2 \ll 1$). Had these approximations, which
are beyond IA, not been made, Eq. (\ref{main}) would be more
complicated. For example, a contamination from the GPD $E_q$ would
be found. Therefore, the formulae we are discussing here are not
strict convolution formulae in the conventional probabilistic
sense. Finally, if $\xi=0$, the formula defining $\tilde H$
becomes a strict convolution (cf. Eq. (5)) and, as a matter of
fact, in this case GPDs have a probabilistic interpretation,
being the Fourier transforms of the impact factor dependent parton
distributions \cite{burk}. Our formalism is therefore consistent
with the fact that convolutions, in the probabilistic sense, arise
once we limit the expressions to quantities having a probabilistic
interpretation.

Let us now specify the formalism to the
proton case and to CQM calculations.
If the $E$-dependence of $\xi'$, i.e.,
the $E$-dependence of $\tilde z$ (cf. Eq. (\ref{xi1}))
is disregarded in Eq. (\ref{spec}), so that
the spin dependent, off-diagonal momentum distribution
\begin{equation}
\tilde n_{q_0}(\vec p, \vec p + \vec \Delta) =
\int dE \tilde P_{q_0}(\vec p, \vec p + \vec \Delta, E )~
\label{clo}
\end{equation}
is recovered, Eq.(\ref{spec}) can be written in the form
\begin{eqnarray}
\tilde H_{q}(x,\xi,\Delta^2) & \simeq &
\sum_{q_0} \int d \vec p
\, \tilde n_{q_0}(\vec p, \vec p + \vec \Delta) {\xi' \over \xi}
\tilde H_{q_0 q}(x',\xi',\Delta^2) = \nonumber
\\
& = &
\sum_{q_0} \int_x^1 {dz \over z} \int d \vec p
\, \tilde n_{q_0}(\vec p, \vec p + \vec \Delta)
\delta \left( z - { \xi \over \xi' }  \right)
\tilde H_{q_0 q} \left( { x \over z }, { \xi \over z },\Delta^2 \right)~,
\label{pnk1}
\end{eqnarray}
which can also be cast in the form of Eq. (\ref{main}),
with the light-cone off-diagonal momentum
distribution given by
\begin{equation}
\tilde H_{q_0}(z, \xi ,\Delta^2 ) =  \int d \vec p
\, \tilde n_{q_0}(\vec p, \vec p + \vec \Delta)
\delta \left( z + \xi  - { p^+ \over \bar P^+ } \right)
\label{hq01}
\end{equation}
which coincides with Eq. (\ref{usu})
and it
is to be evaluated in a
given CQM,
for $q_0=u_0$
or $d_0$, while $\tilde H_{q_0 q}( {x \over z},
{\xi \over z},\Delta^2)$ is the constituent quark
helicity-dependent GPD, which is still to be discussed and
will be modelled in the next section.

\section{A model for the helicity-dependent GPD of the Constituent Quark}

The main issue remaining is the definition of $\tilde H_{q_0 q}(
{x \over z}, {\xi \over z},\Delta^2)$, the helicity-dependent
constituent quark GPD, appearing in Eq. (\ref{main}).

As it has been done in Ref. \cite{prd} for the unpolarized case,
this quantity will be modelled to respect that its forward limit,
recovers the constituent quark parton distributions. We adopt the
picture proposed in \cite{scopetta1,scopetta2} that the
constituent quark is a complex system of point-like partons
described by the scheme suggested by Altarelli, Cabibbo, Maiani
and Petronzio (ACMP) \cite{acmp}. In this scheme the constituent
quarks are composite objects whose structure functions are
described by a set of functions $\phi_{q_0q}(x)$ that specify the
number of point-like partons of type $q$ which are present in the
constituent of type $q_0$, with fraction $x$ of the total
momentum. We call these functions, generically, the structure
functions of the constituent quark.

The functions describing the nucleon parton distributions are expressed in
terms of the independent $\phi_{q_0q}(x)$ and of the constituent
density distributions ($q_0=u_0,d_0$) as,
\bq
q(x,Q^2)=\sum_{q_0}\int_x^1 {dz\over z}
q_0(z,Q^2) \phi_{q_0q} \left({x \over z},Q^2 \right)~,
\label{fx}
\eq
where $q$ labels the various partons, i.e., valence quarks
($ u_v,d_v$), sea quarks ($u_s,d_s, s$), sea antiquarks ($\bar u,\bar d, \
\bar
s$) and gluons $g$.

The different structure functions of the
constituent quarks are derived from three very natural assumptions
\cite{acmp}:
i) the point-like partons are $QCD$ degrees of freedom, i.e.
quarks, antiquarks and gluons;
ii) Regge behavior for $x\rightarrow 0$ and duality ideas;
iii) invariance under charge conjugation and isospin.

These considerations define in the case of the valence quarks the following
structure function

\bq
\phi_{qq_v}({x},Q^2)
= { \Gamma(A + {1 \over 2}) \over
\Gamma({1 \over 2}) \Gamma(A) }
{ (1-x)^{A-1} \over \sqrt{x} }.
\label{csf1}\eq
For the sea quarks the corresponding structure function becomes,

\bq
\phi_{qq_s}({x},Q^2)
= { C \over x } (1-x)^{D-1},\label{csf2}
\eq
and, in the case of the gluons, it is taken

\bq
\phi_{qg}({x},Q^2)
= { G \over x } (1-x)^{B-1}~.\label{csf3}
\eq

The last input of the approach consists of the definition of the
scale at which the constituent quark structure is defined. We
choose the so called hadronic scale $\mu_0^2$ \cite{grv,trvv}.
This last hypothesis fixes $all$ the parameters of the approach
(Eqs. (\ref{csf1}) through (\ref{csf3})). The constants $A$, $B$,
$G$ and the ratio $C/D$ are determined by the amount of momentum
carried by the different partons, corresponding to a hadronic
scale of $\mu_0^2=0.34$ GeV$^2$, according to the parameterization
of \cite{grv}. $C$ (or $D$) is fixed according to the value of
$F_2$ at $x=0.01$ \cite{acmp}, and its value is chosen again
according to \cite{grv}. We stress that all these inputs are
forced only by the updated phenomenology, through the 2$^{nd}$
moments of PDFs. The values of the parameters obtained are listed
in \cite{scopetta1}.

The other ingredients appearing in Eq. ({\ref{fx}}), i.e.,
the density distributions for
each constituent quark, represent the unpolarized equivalent of
Eq. (\ref{usu}).

In Ref. \cite{scopetta2}, the spin dependent
structure function of the constituent quark have been
built generalizing the analysis of Ref. \cite{scopetta1},
leading to Eqs. (\ref{csf1}) -- (\ref{csf3}).
In here, we review the
procedure addressed in Ref. \cite{scopetta2}.
Let

\bq
\Delta q (x,\mu_0^2) = q_+ (x,\mu_0^2) - q_- (x,\mu_0^2)
\eq
where $\pm$ label the quark spin projections and $q$ represents any flavor.
The $ACMP$ approach,
generalized to the polarized case,
 implies
\bq
q_i(x) = \sum_{q_0} \int_x^1 \frac{dz}{z} \sum_j q_{0j}(z)
\Phi_{q_{0j}q_i} (\frac{x}{z})
\eq
where $i=\pm$ labels the partonic spin projections, $j=\pm$ the constituent
quark spins and $q_0=u_0,d_0$. Using spin symmetry we arrive at

\bq
\Delta q (x) =\int_x^1 \frac{dz}{z} \left [ \Delta u_0 (z)
\Delta \Phi_{u_0q}(\frac{x}{z})
+ \Delta d_0 (z)\Delta \Phi_{d_0q}(\frac{x}{z})) \right ]
\eq
where $\Delta q_0 = q_{0+} - q_{0-}$, and

\bq
\Delta \Phi_{q_0q} = \Phi_{q_0+q+} - \Phi_{q_0+q-}
\eq
Note at this point that the unpolarized case can be described in this
generalized formalism as

\bq
 q (x) =q_+(x) + q_-(x) =
\sum_{q_0} \int_x^1 \frac{dz}{z} q_0 (z) \Phi_{q_0q} (\frac{x}{z})~,
\eq
where

\bq
\Phi_{q_0q} = \Phi_{q_0+q+} + \Phi_{q_0+q-},
\eq

The description, reformulated in terms of the conventional valence and sea
quark separation, yields

$$\Delta q (x) =\Delta q_v (x) +\Delta q_s (x) = $$
\bq
\sum_{q_0}
\int_x^1 \frac{dz}{z} \Delta q_0 (z)
\left [ \Delta \Phi_{q_0 q_v}(\frac{x}{z})
+\Delta \Phi_{q_0 q_s}(\frac{x}{z}) \right ]
\label{dfx}
\eq
Since $\Delta \Phi_{{q_0} q_v} = \Delta \Phi_{q_0 q_v} \delta_{{q_0} q}$,
i.e. any constituent has a leading parton with
the same quantum numbers, one has
\bq
\Delta q_v (x) =
\int_x^1 \frac{dz}{z} \Delta q_0 (z) \Delta \Phi_{q_0q_v}
(\frac{x}{z}),
\eq
\bq
\Delta q_s (x) =
\sum_{q_0}
\int_x^1 \frac{dz}{z}
\Delta q_0 (z) \Delta \Phi_{q_0q_s}
(\frac{x}{z})
\label{sea}
\eq
Introducing $SU(6)$ (spin-isospin) symmetry as a simplifying assumption, gives

\bq
\Delta \Phi_{u_0u} = \Delta \Phi_{d_0d}
\eq
and

\bq
\Delta \Phi_{u_0d} = \Delta \Phi_{d_0u}.
\eq
 Furthermore they imply

\bq
\Delta \Phi_{u_0 u_v} +  \Delta \Phi_{u_0 u_s} =
\Delta \Phi_{d_0 d_v} +  \Delta
\Phi_{d_0 d_s}
\eq
and

\bq
\Delta \Phi_{u_0 d_s} = \Delta \Phi_{d_0 u_s}.
\eq

By adding  to these the following relations

\bq
\Delta \Phi_{u_0 u_s} = \Delta \Phi_{d_0 u_s},
\label{simp1}
\eq

\bq \Delta \Phi_{d_0 d_s} = \Delta \Phi_{u_0 d_s}, \label{simp2}
\eq which are beyond $SU(6)$ symmetry, but quite reasonable, one
obtains,

\bq
\Delta \Phi_{u_0 u_s} = \Delta \Phi_{d_0 u_s} =
\Delta \Phi_{u_0 d_s} =
\Delta \Phi_{d_0 d_s} = \Delta \Phi_{q_0 q_s}
\label{phisea}
\eq
and

\bq
\Delta \Phi_{u_0 u_v} =\Delta \Phi_{d_0 d_v} =\Delta \Phi_{q_0 q_v}.
\eq

In this way the structure functions for the valence
and for the sea are reduced to
just two independent constituent structure functions and Eq. (\ref{sea})
simplifies to

\bq
\Delta q_s (x) = \int_x^1 \frac{dz}{z} (\Delta u_0 (z)  + \Delta d_0 (z))
\Delta \Phi_{q_0q_s}(\frac{x}{z}).
\label{deltasea}
\eq

The same argumentation applied to gluons implies

\bq
\Delta g (x) = \int_x^1 \frac{dz}{z} (\Delta u_0 (z) + \Delta d_0 (z))
\Delta \Phi_{q_0 g}(\frac{x}{z})
\eq

Thus the $ACMP$ procedure can be extended to the polarized case by
introducing three additional structure functions for the
constituent quarks: $\Delta \Phi_{q_0 q_v}$, $\Delta \Phi_{q_0
q_s}$ and $\Delta \Phi_{q_0 g}$.

In order to determine the polarized constituent structure
functions we add some assumptions which will determine them and
those of the unpolarized case using a small number of parameters.
They are: iv) factorization assumption: $\Delta \Phi$ cannot
depend upon the quark model used, i.e, cannot depend upon the
particular $\Delta q_0$; v) positivity assumption: the positivity
constraint $\Delta \Phi \leq \Phi $ is saturated for $x = 1$.
These additional assumptions determine completely the parameters
of the polarized constituent structure functions. In fact, the
$QCD$ partonic picture, Regge behavior and duality imply that

\bq
\Delta\Phi_{q_0 f} = {\Delta C_f
{x^{a_f}}
(1-x)^{A_f - 1}}
\eq
and
$ 0 < a_f < \frac{1}{2}$,
for all $f= q_v, q_s, g$, as defined by dominant
exchange of the $A_1$ meson trajectory \cite{Heimann-Ellis}.

The positivity restriction, $\Delta \Phi \leq \Phi$, is a natural
consequence of the probability interpretation of the parton
distributions. The assumption that the inequality is saturated for
$x=1$, in the spirit of ref. \cite{kaur}, implies that $\Delta C_f
= C_f$, the latter being the parameter fixed in the analysis of
the unpolarized case, and therefore when $x \approx 1$ the partons
which carry all of the momentum also carry all of the
polarization, i.e., $\Phi_{+-} = 0$. As is discussed in Ref.
\cite{scopetta2}, this scheme has been changed only for the sea
structure function, in order to obtain a reasonable description of
the polarized sea. This led to the definition of a parameter,
$\Delta C_{q_s}$, different from the one, $C_{q_s}$ defined in the
unpolarized case. From the point of view of the number of
parameters, this is a minimal assumption, since it reduces the
parameters of the polarized case to those of the unpolarized one,
with the only exception of $C_{q_s}$. Lastly the exponents $A_f$
are also taken from the unpolarized case. To summarize the
parameterization, let us stress that the main change between the
polarized functions and the unpolarized ones comes only from Regge
behavior and the only parameter which has been changed is the one
related to the polarized sea distribution.

With all the above hypothesis the constituent quark functions become

\bq
\Delta \Phi_{qq_v}({x}, \mu_0^2)
= { \Gamma(A + {1 \over 2}) \over
\Gamma({1 \over 2}) \Gamma(A) } x^{a_f}
{ (1-x)^{A-1} }
\label{pi}
\eq
\bq
\Delta \Phi_{qq_s}({x}, \mu_0^2)
= \Delta C_s  x^{a_f}
{ (1-x)^{D-1} }
\eq
\bq
\Delta \Phi_{qg}({x}, \mu_0^2)
= G  x^{a_f}
{ (1-x)^{B-1} }
\label{pf}
\eq
where $A,D,B,G$ are the parameters of the unpolarized case,
while $\Delta C_s$ is taken to reproduce the
polarized data. In what follows,
we shall use
$0 \leq a_f \leq0.5$, the range proposed in Ref. \cite{abfr}
in agreement with Ref. \cite{Heimann-Ellis}.

We have to generalize this scenario to describe off-forward
phenomena. Of course, the forward limit of our GPDs formula, Eq.
(\ref{main}), has to be given by Eq. ({\ref{dfx}}). By taking the
forward limit of Eq. (\ref{main}), one obtains:

\begin{eqnarray}
\tilde H_{q}(x,0,0) & = &
\sum_{q_0} \int_x^1 { dz \over z}
\tilde H_{q_0}(z, 0, 0 )
\tilde H_{q_0 q} \left( {x \over z},0,0 \right)
\nonumber \\
& = &
\sum_{q_0} \int_x^1 { dz \over z}
\Delta q_0(z)
\tilde H_{q_0 q} \left( {x \over z},0,0 \right)~,
\label{for}
\end{eqnarray}

so that, in order for the latter to coincide with Eq.
(\ref{dfx}), one must have
$\tilde H_{q_0 q}( x,0,0) \equiv
\Delta \phi_{q_0q}(x)$.

In such a way, through the ACMP prescription, the forward
limit of the unknown constituent quark GPD
$\tilde H_{q_0 q}( {x \over z},
{\xi \over z},\Delta^2)$ can be fixed.

The off-forward behavior of the Constituent Quark GPDs has to be
modelled and it will be done, as in Ref. \cite{prd} for the
unpolarized case, using the ``$\alpha$-Double Distributions''
(DD's) language proposed by Radyushkin \cite{radd,rag}. The DD's,
$\Phi(\tilde x, \alpha, \Delta^2)$, are a representation of GPDs
which automatically guarantees the polynomiality property. GPDs
can be obtained from DD's after a proper integration, and, as is
explained in \cite{radd,rag}, parton densities are recovered in
the forward limit, while meson distribution amplitudes are
obtained in the $P=0$ limit of the DD's. In some cases, such a
transparent physical interpretation, together with the symmetry
properties which are typical of distribution amplitudes ($\alpha
\rightarrow - \alpha$ symmetry), allows a direct modelling,
already developed in \cite{rad1}.

The relation between any GPD $\tilde H$, defined {\sl \`a la Ji},
for example the one we need, i.e.
$\tilde H_{q_0q}$ for the constituent quark target,
is related to the $\alpha$-DD's, which we call
$\Delta \tilde \Phi_{q_0 q} (\tilde x, \alpha,\Delta^2)$ for the constituent
quark,
in the following way
\cite{radd,rag}:
\begin{eqnarray}
\tilde H_{q_0q}(x,\xi,\Delta^2) = \int_{-1}^1 d\tilde x
\int_{-1 + |\tilde x|}^{1-|\tilde x|}
\delta(\tilde x + \xi  \alpha - x)
\Delta \tilde \Phi_{q_0 q} (\tilde x, \alpha,\Delta^2) d \alpha~.
\label{hdd}
\end{eqnarray}

With some care, the expression above can be integrated over
$\tilde x$ and the result is explicitly given in \cite{rag},
where
a factorized ansatz is suggested for the DD's which, for the
helicity-dependent one, could be written:
\begin{equation}
\Delta \tilde \Phi_{q_0 q} (\tilde x, \alpha,\Delta^2) =
h(\tilde x, \alpha)
\Delta \Phi_{q_0 q} (\tilde x)
G_{q_0}^q(\Delta^2)~,
\label{ans}
\end{equation}
with the $\alpha$ dependent term, $h(\tilde x, \alpha )$, having
the character of a mesonic amplitude.

Besides, in Eq. (\ref{ans})
$
\Delta \Phi_{q_0 q} (\tilde x)
$
represents the forward density
and, eventually,
$
G_{q_0}(\Delta^2)
$
the constituent quark axial ff.

One immediately realizes that the helicity-dependent GPD of the
constituent quark, Eq. (\ref{hdd}), with the factorized form Eq.
(\ref{ans}), fulfills the crucial constraints of GPDs, i.e., the
forward limit, the first-moment and the polynomiality condition,
the latter being automatically verified in the DD's description.

In the following, it will be assumed for the constituent quark GPD the
above factorized form, so that one needs to model the three functions
appearing in Eq. (\ref{ans}), according to the present
description.

For the amplitude $h$, use will be made of
one of the simple normalized forms suggested in
\cite{radd},
on the bases of the symmetry
properties of DD's:
\begin{equation}
h(\tilde x, \alpha) = {3 \over 4} { (1 - \tilde x)^2 - \alpha^2
\over (1 - \tilde x)^3 }~.
\label{h}
\end{equation}

In addition, since
we will identify
quarks for $x \ge \xi/2$, pairs for $ x \le |\xi/2|$,
antiquarks for  $x \le -\xi/2$, and, since
in our approach the forward densities
$\Delta \Phi_{q_0q} (\tilde x)$ have to be given by
the standard $\Delta \Phi$ functions of the
$ACMP$ approach, Eqs. (\ref{pi})--(\ref{pf}),
one has, for the DD of flavor $q$
of the constituent quark:
\begin{eqnarray}\label{sr}
\Delta \tilde \Phi_{q_0 q} (\tilde x, \alpha,\Delta^2) =
\cases{
(h(\tilde x,\alpha) \tilde \Phi_{q_0q_v}(\tilde x)
+
h(\tilde x,\alpha) \tilde \Phi_{q_0q_s}(\tilde x)) G_{q_0}^q(\Delta^2)
 &
for $\tilde x \ge 0$ \cr
- h(- \tilde x,\alpha) \tilde \Phi_{q_0q_s}(- \tilde x)
G_{q_0}^q(\Delta^2)
 & for $\tilde x < 0$ \cr}
\label{dd}
\end{eqnarray}

The above definition,
when integrated over $\alpha$
gives the correct limits \cite{rag}:
\begin{equation}
\int_{-1 + \tilde x}^{1-\tilde x} \Delta \tilde \Phi_{q_0q}
(\tilde x, \alpha,\Delta^2=0)|_{\tilde x > 0}
d \alpha = \Delta \Phi_{q_0q} (\tilde x)~,
\end{equation}
and
\begin{equation}
\int_{-1 + |\tilde x|}^{1-|\tilde x|} \Delta \tilde \Phi_{q_0q}
(\tilde x, \alpha,\Delta^2=0)|_{\tilde x < 0} d \alpha =
- \Delta \Phi_{q_0 \bar q}
(-\tilde x)~.
\end{equation}

Eventually, as axial ff we will take a simple monopole form
corresponding to a constituent quark size $r_Q \simeq 0.3 fm$:

\begin{equation}
G_{q_0}(\Delta^2) = { 1 \over 1 - {\Delta^2 r_Q^2
\over 6 }}~,
\label{ffacmp}
\end{equation}

showing i.e. the same $Q^2$ dependence of the electromagnetic ff
used in \cite{prd},
the latter scenario being strongly supported by the
analysis of \cite{psr}.

By using such a ff and Eq. (\ref{h}), together with the standard
ACMP $\Delta \Phi$'s, Eqs. (\ref{pi}) --
(\ref{pf}), in Eq. (\ref{dd}), and inserting the obtained
$\Delta \Phi_{q_0 q} (\tilde x, \alpha,\Delta^2)$
into Eq. (\ref{hdd}),
the constituent quark GPD in the ACMP scenario can be eventually calculated.

Let us discuss our factorized choice for the DD of the constituent
quark, Eq. (\ref{ans}). Such a factorization does not hold for the
GPD of the proton \cite{md}. No microscopic calculation supports
it  \cite{epj}. The present approach is consistent with the
absence of factorization of the {\sl{proton}} GPD in
$\Delta^2-$dependent and $\Delta^2-$independent terms since an
additional, crucial, $\Delta^2$ dependence comes from the quantity
$H_{q_0}$ in Eq. (\ref{main}), i.e., from the proton wave function
of the quark model used.

The factorized form is assumed for the GPD of the constituent
quark for simplicity. We do not know what the helicity-dependent
GPD of the constituent quark is and we start modelling it by using
the simplest possible structure which fulfills the required
constraints. Clearly this choice deserves a deeper explanation,
which we do not have at the moment. However, one expects that the
structure of the constituent quark GPD is closer to the asymptotic
behavior, than that of the proton, because it probes higher
energies. The non perturbative contributions arise in the precise
shape of the three functions building the GPD, which we take from
general QCD based arguments and from data. It is true that even
the forward structure of the constituent quark is not known, but
we use a parameterization which performs well in describing the
data \cite{scopetta2}. In this sense, the forward limit is well
reproduced, as shown in the next section. On the contrary, the
profile Eq. (\ref{h}) is {\sl{assumed}} to be valid also for the
constituent quark, even in the helicity dependent case, simply
because it satisfies some of the required properties.

These arguments and a small set of parameters, fixed once forever,
make our approach able, if a proper CQM is used, to describe DIS
data for parton distributions, elastic form factors, and to
predict unpolarized and polarized GPDs. It allows for an effective
description of the known structure of the nucleon and its
predictive power needs to be tested with new observables.

\section{Results and discussion}

In this section we present the results obtained
for the helicity-dependent GPD $\tilde H_q$ of the proton,
for $\xi^2 \ll 1$ and $\vec \Delta^2 \ll m^2$,
according to the approach described so far.
The main equation to be evaluated is
Eq. (\ref{main}), written again here below for the sake of clarity:
\begin{eqnarray}
\tilde H_{q}(x,\xi,\Delta^2) =
\sum_{q_0} \int_x^1 { dz \over z}
\tilde H_{q_0}(z, \xi ,\Delta^2 )
\tilde H_{q_0 q}\left( {x \over z},
{\xi \over z},\Delta^2 \right)~.
\nonumber
\end{eqnarray}

In the above equation,
the quantity $\tilde H_{q_0q}$, the helicity-dependent
constituent quark GPD, is modelled
according to the arguments described in the previous section.
This means that it is obtained evaluating Eq. (\ref{hdd}),
where the DD of the constituent quark,
$\Delta \tilde \Phi_{q_0 q} (\tilde x, \alpha,\Delta^2)$ ,
is given by Eq. (\ref{dd}), calculated in turn through
the ff, Eq. (\ref{ffacmp}), the function
$h$,  Eq. (\ref{h}), together with the forward
ACMP $\Delta \Phi$'s, Eqs. (\ref{pi}) -- (\ref{pf}).

The other ingredient in Eq. (\ref{main}),
$\tilde H_{q_0}$, has been evaluated according to
Eq. (\ref{usu}):
\begin{eqnarray}
\tilde H_{q_0}(z, \xi ,\Delta^2 ) =  \int d \vec p
\,
\tilde n_{q_0}(\vec p, \vec p + \vec \Delta)
\delta \left( z + \xi  - { p^+ \over \bar P^+ } \right)~.
\nonumber
\end{eqnarray}
The calculation has been performed in the Breit frame,
where one has, in the NR limit studied, $\sqrt{2} \bar P^+ \rightarrow M$.
The off-diagonal momentum distribution appearing in the
formula above,
$\tilde n_{q_0}(\vec p, \vec p + \vec \Delta)$, defined in
Eq. (\ref{clo}), has been evaluated within
the Isgur and Karl (IK) model \cite{ik}, according
to the definition
\be
\tilde n_{q_0}(\vec p, \vec p + \vec \Delta) =
\tilde n_{q_0+}(\vec p, \vec p + \vec \Delta)
-
\tilde n_{q_0-}(\vec p, \vec p + \vec \Delta)~,
\eq

where $n_{q_{0\pm}} (\vec p, \vec p + \vec \Delta)$ is the
off-diagonal
momentum distribution for the valence quark momentum $\vec{p}$
and spin projection
parallel or antiparallel to that of the nucleon.
$n_{q\pm} (\vec{p},\vec p + \vec \Delta)$ can be evaluated
projecting out the appropriate spin and flavor component of the constituent
quark and in the corresponding baryonic state is given by

\bq
\tilde n_{u_0/d_0 , \pm}
(\vec p, \vec p + \vec \Delta)
& = & \;3\;\left ( _{sf}<N ,\; J_z = +\frac{1}{2}|
\frac{1 \pm \tau^z_3}{2}\frac{1 \pm\sigma^z_3}{2}|N ,\; J_z = +\frac{1}{2}>
_{sf} \right ) \times
\nonumber
\\
& \times &
\int d \vec p_1 d \vec p_2 \psi^*(\vec p_1,\vec p_2,\vec
p + \vec \Delta) \psi(\vec p_1,\vec p_2,\vec p )~,
\eq
where $ |N ,\; J_z = +\frac{1}{2}>
_{sf}$ is the spin-flavor state and $\psi(\vec p_1,\vec p_2,\vec p_3)$
is the proton wave function in momentum space
(cf. Ref. \cite{epj} for the definition
in the unpolarized case).

In the IK CQM \cite{ik},
including contributions up to the $2 \hbar \omega$ shell,
the proton state is given by the
following admixture of states
\begin{eqnarray}
|N \rangle =
a_{\cal S} | ^2 S_{1/2} \rangle_S +
a_{\cal S'} | ^2 S'_{1/2} \rangle_{S} +
a_{\cal M} | ^2 S_{1/2} \rangle_M +
a_{\cal D} | ^4 D_{1/2} \rangle_M~,
\label{ikwf}
\end{eqnarray}
where the spectroscopic notation $|^{2S+1}X_J \rangle_t$,
with $t=A,M,S$ being the symmetry type, has been used.
The coefficients were determined by spectroscopic properties to be
\cite{mmg}:
$a_{\cal S} = 0.931$,
$a_{\cal S'} = -0.274$,
$a_{\cal M} = -0.233$, $a_{\cal D} = -0.067$.
\\
The results for the
GPD $\tilde H(x,\xi,\Delta^2)$,
neglecting in (\ref{ikwf})
the small $D$-wave contribution, are found to be:

\begin{eqnarray}
\tilde H_u(x,\xi,\Delta^2)
& = & 3 { M \over \alpha^3}
\left( {3 \over
2 \pi } \right)^{3/2}
e^{ { \Delta^2 \over 3 \alpha^2 }}
\int dk_x \int dk_y \,
f_0(k_x,k_y,x,\xi,\Delta^2) \nonumber
\\
& \times &
\left [ \Delta f_s(k_x,k_y,x,\xi,\Delta^2)
+ \Delta {\tilde f} (k_x,k_y,x,\xi,\Delta^2) \right ]~,
\label{iku}
\end{eqnarray}

\begin{eqnarray}
\tilde H_d(x,\xi,\Delta^2)
& = & 3 { M \over \alpha^3}
\left( {3 \over
2 \pi } \right)^{3/2}
e^{{ \Delta^2 \over 3 \alpha^2 }}
\int dk_x \int dk_y \,
f_0(k_x,k_y,x,\xi,\Delta^2) \nonumber
\\
& \times &
\left [ -{1 \over 4} \Delta f_s(k_x,k_y,x,\xi,\Delta^2)
+ {1 \over 5} \Delta {\tilde f} (k_x,k_y,x,\xi,\Delta^2)
+ \bar f(k_x,k_y,x,\xi,\Delta^2)
\right ]~,
\label{ikd}
\end{eqnarray}

for the flavors $u$ and $d$, respectively,
with

\begin{eqnarray}
f_0(k_x,k_y,x,\xi,\Delta^2) =
{ \bar k_0 \over \bar k_0 + \bar k_z }
f_\alpha(\Delta_x,k_x)
f_\alpha(\Delta_y,k_y) f_\alpha(\Delta_z,\bar k_z)~,
\end{eqnarray}
\begin{eqnarray}
f_\alpha(\Delta_i,k_i) = e^{ - {1 \over \alpha^2} \left( {3 \over 2} k_i^2
+ k_i\Delta_i \right) }~~,
\end{eqnarray}
\begin{eqnarray}
\bar{k_z} = { M^2(x + \xi)^2 - ( m^2 + k_x^2 + k_y^2)
\over 2 M ( x + \xi) }~,
\end{eqnarray}

\begin{eqnarray}
\label{fs}
\Delta f_s(k_x,k_y,x,\xi,\Delta^2) & = & {4 \over 9} a_s^2 +
{4 \over 9} a_{s'}^2
\left[ {5 \over 4} - {3 \over 2}{k^2 \over \alpha^2} + { 3 \over 4}
{k^4 \over \alpha^4} + { 1 \over {\alpha^2}}
\left ( - {\Delta^2 \over 3} + \vec \Delta \cdot \vec k \right )
\left ( {k^2 \over \alpha^2} -1 \right ) \right ] \nonumber
\\
& + &
{4 \over 9 } a_M^2 \left [ {5 \over 8} - {3 \over 4} {k^2 \over \alpha^2}
+ { 3 \over 8} {k^4 \over \alpha^4}
{ 1 \over {2 \alpha^2}}
\left ( -{\Delta^2 \over 3} + \vec \Delta \cdot \vec k \right )
\left ( {k^2 \over \alpha^2} -1 \right ) \right ]
\\
& - & { 4 \over 9} \sqrt{3}  a_S a_{S'}
\left [ \left ( - 1 + {k^2 \over \alpha^2} \right )
- { 2 \over {3 \alpha^2}}
\left ( {\Delta^2 \over 3} - \vec \Delta \cdot \vec k \right )
\right ]~, \nonumber
\end{eqnarray}

\begin{eqnarray}
\label{ft}
\Delta \tilde f(k_x,k_y,x,\xi,\Delta^2) & = &
- { 1 \over \sqrt{2}}
{5 \over 3} a_M a_{S'}
\left [ { 1 \over 6} - {k^2 \over \alpha^2} +
{1 \over 2} {k^4 \over \alpha^4}
- { 2 \over {3 \alpha^2}}
\left ( -{\Delta^2 \over 3} - \vec \Delta \cdot \vec k \right )
\right .
\nonumber
\\
& + & \left . { 2 k^2 \over {3 \alpha^4}}
\left ( -{\Delta^2 \over 3} + \vec \Delta \cdot \vec k \right )
\right ]~,
\end{eqnarray}

\begin{eqnarray}
\label{bf}
\bar f(k_x,k_y,x,\xi,\Delta^2)  =
{ 1 \over 9} {k \over \ \alpha^2} \sqrt{ { 9 \over 4} k^2 -
\Delta^2 + 3 \vec \Delta \cdot \vec k }~,
\end{eqnarray}

and $\bar k_0 = \sqrt{ m^2+k_x^2+k_y^2+\bar k_z^2}$, being $m \simeq
M/3$ the constituent quark mass.
Here we have used the notation $k^2=\vec k^2, \vec k = (k_x,k_y,k_z)$.
Evaluating the forward limit, $\Delta^2=0, \xi =0$,
of Eqs. (\ref{iku}) and (\ref{ikd}),
the expressions of
the polarized quark densities in the IK model,
obtained in \cite{rope}, are consistently recovered.

The harmonic oscillator parameter,
$\alpha$, of the IK model,
has ben chosen so that the experimental
r.m.s. of the proton is reproduced by the slope,
at $\Delta^2=0$,
of the charge ff, so that
a value of $\alpha=1.18 f^{-1}$ is obtained
\cite{prd}.
Such a ff reproduces well the data at the low values of
$\Delta^2$ which are accessible in the present approach.
For higher values of $\Delta^2$, it would not be realistic
\cite{mmg}.
The same quality of results is expected for the proton axial
form factor, which is obtained
by integrating Eq. (\ref{main}) over $x$, according
to Eq. (\ref{ffc}). The experimental dipole trend
\cite{mainz}
is qualitatively
recovered at the low values of $\Delta^2$ under scrutiny
here.

%

Results for the $u$-quark $\tilde H_u$ distribution, at the scale of the model
$\mu_o^2$,
are shown in Figs 2 to 4.
In Fig. 2, it is shown
for $\Delta^2=-0.1$ GeV$^2$ and $\xi=0.1$.
One should remember that the present approach does not allow
to estimate realistically the region $-\Delta^2 \ge m^2 \simeq 0.1$ GeV$^2$,
so that we are showing here the result corresponding
to the highest possible $\Delta^2$ value.
Accordingly, the maximum value of
the skewedness
is therefore
$\xi \simeq 0.17$ (cf. Eq. (\ref{xim})), fulfilling
the requirement $\xi^2 \ll 1$.
The dashed curve represents what is obtained in the pure
IK model, i.e., by evaluating Eq.
(\ref{iku}). One should notice that such a result, obtained
in a pure valence CQM, should
vanish for $x \le \xi$ and for $x \ge 1$. The small tails
which are found in these forbidden regions represent the amount
of support violation of the approach.
In particular, for the shown values of $\Delta^2$ and $\xi$, a violation
of 2 $\%$ is found. In general, in the accessible region the violation
is never larger than few percents.
The full curves in Fig. 3 represents the complete result of the present
approach, i.e., the evaluation of Eq. (\ref{main}) following the steps
and using the ingredients described in this section and in the
previous one.
The two curves are obtained in correspondence
of the two extreme values of the Regge intercept
\cite{Heimann-Ellis},
$a_f=0$ (upper curve) and $a_f=0.5$
(lower curve), used to define
the helicity dependent constituent quark
structure functions \cite{scopetta2}.
A relevant
contribution is found to lie in the ERBL region,
in agreement with other estimates \cite{pog}.
As already explained, the knowledge
of GPDs in the ERBL region is a crucial prerequisite for
the calculation of all the cross-sections and the observables
measured in the processes where GPDs contribute.
We notice that the ERBL region is accessed here, with
respect to the approach of Ref. \cite{epj} generalized
to the helicity-dependent case,
which gives the dashed curve, thanks
to the constituent structure which has been introduced
implementing the ACMP procedure.
A similar result has been found in \cite{prd} in the
unpolarized case.

In Fig. 3, special emphasis is devoted to show the
$\xi$-dependence of the results. For the allowed
$\xi$ values, $\tilde H_u(x,\xi,\Delta^2)$,
evaluated using our main formula, Eq. (\ref{main}),
is shown for
four different values of $x$. It is clearly seen that such
a dependence is stronger in the ERBL region,
than in the DGLAP region, in good agreement
with other estimates \cite{pog} and
with our previous analysis, performed in the unpolarized case
\cite{prd}.
To allow for a complete view of the outcome of our approach,
in Fig. 4 the $x$ and $\xi$ dependences are shown together
in a 3-dimensional plot.

The results shown so far are associated with
the low scale of the model, the hadronic scale $\mu_0^2$,
fixed to the value 0.34 GeV$^2$, as discussed in Section 4.

In order to have a feeling of the quality
of the present predictions,
since no data are available at the moment for $\tilde H(x, \xi, \Delta^2)$,
we evaluate in the following the forward limit of the obtained quantities.
In Figs. 5 and 6,
the proton and neutron spin dependent
structure function $g_1(x,Q^2)$,
built from the forward limit of Eq. (\ref{main})
evaluated for $q=q_v,q_s,g$,
are compared with experimental data.
In Ref. \cite{scopetta2}, the succesful analysis had been performed using
the model of Ref. \cite{iac}, so that the encouraging results
shown there have to be confirmed when the IK model is considered,
as it is being done here.
To evaluate $g_1(x,Q^2)$ at the experimental scale,
a NLO evolution of the model parton distributions
has been performed. It is known that perturbative $QCD$ to this
order allows the proposal of varied factorization schemes \cite{fp},
whereby one is able to define the partonic distributions in
different ways without altering the physical observables. In our analysis the
parton distributions are determined by the quark model through the ACMP
procedure. It is evident that
different factorization schemes lead to different results.
Following Ref. \cite{scopetta2},
we have adopted on
physical grounds the $AB$ scheme as defined in ref.\cite{bfr} \footnote{It
consists in modifying minimally the NLO polarized anomalous dimensions
\cite{NLOP}, calculated in the $\overline{MS}$ scheme,
in order to have the axial anomaly governing the first moment of $g_1$,
as proposed in Ref \cite{Altarelli-Ross}.}. In it some relevant physical
observables, such as $\Delta \Sigma$, the total spin
carried by the quarks,
are scale independent, i.e., they behave as conserved quantities,
and therefore the partons have a well defined interpretation in terms of
constituents.
Fig. 5 and 6 show that the present approach describes well
the data \cite{emc,smc} for the proton structure function $g_1$, and
the trend of the neutron data \cite{e154} is also reasonably reproduced.
The quality of the agreement is comparable with that
obtained in Ref. \cite{scopetta2} using the model of \cite{iac}.

The present approach in the forward limit, i.e.
the description given in Ref. \cite{scopetta2}, also
allows for the predictions of axial charges of the proton.
By using the IK model, the following values are obtained
(the first listed value corresponds to the Regge slope
$a_f = 0.5$, the second to $a_f=0.$):
for the octet charge, $a_8 = 0.540  \div 0.671$
(experimental value: $0.58 \pm 0.03$ \cite{cr}),
for the isospin charge, $a_3 = 0.857  \div 1.063$
(experimental value: $1.257 \pm 0.003$ \cite{pdg}),
while the predictions for the
singlet charge and for the spin carried by the gluons
lead to
$\Delta \Sigma = 0.402 \div 0.530$
(from the analysis of Ref. \cite{abfr} : $0.45 \pm 0.09$).
Again, the agreement with data of these results, obtained
by using the IK wave functions,
is comparable to the one obtained in Ref. \cite{scopetta2}
by using the wave functions of Ref. \cite{iac}.

This finding deserves a further comment.
The lack of high momentum components
in the IK wave function with respect to that of the
model of Ref. \cite{iac} would indicate the latter as a better
tool to describe the data.
Nevertheless, the low-$x$ behavior of the model results
turns out to be mainly governed by the constituent quark structure
rather than by the constituent quark wave function.
The correct Regge-behavior given by our parametrization
of the constituent quark structure functions
provides us with a model independent prediction.
Such a feature has not to do with the use of a relativistic model,
which would improve only the high $x$ behavior.
As discussed in Ref \cite{nsv}, relativity and constituent quark
structure are independent features, both
of them necessary
if model calculations are used
for a proper description of data in the full $x$ range.

Since no data are presently available for helicity-dependent
GPDs, it is certainly useful to compare the results of model
calculations with model independent constraints, if available.
It has been proposed that they could be
obtained from analysis of data in particular kinematical
conditions. Very recently,
this issue has been addressed
in two papers, where GPDs have been obtained, at $\xi=0$,
by fitting the first moment of a reasonable ansatz
of GPDs to ff data \cite{gpdff1,gpdff2}.
Anyway, such a procedure is difficult to be realized for
the quantity under scrutiny here, due to the lack of a satisfactory
experimental knowledge of the axial ff, which cannot provide
us with a clear constraint on $\tilde H_q(x,0,\Delta^2)$.
As a matter of fact, in Ref. \cite{gpdff2} no prediction has been
produced for it, so that no comparison is possible between
the present results and those of \cite{gpdff2}.
On the contrary, some curves reproducing $\tilde H_q(x,0,\Delta^2)$
have been shown in \cite{gpdff1}.
Nevertheless, as the authors of that paper clearly state,
the lack of precise data for the axial ff does not
allow a realistic parameterization of
$\tilde H_q(x,0,\Delta^2)$,
but rather a simple ansatz for it.
This ansatz is given by the product
of the polarized parton distributions, taken from
the ``scenario 1'' of Ref. \cite{parpol}
at $Q^2 = 4 $ GeV$^2$, and an exponential term,
giving the $\Delta^2$ dependence and additional
$x$ dependence, taking into account the different Regge
behavior at different values of the momentum transfer.
The latter dependence is borrowed from the
analysis performed for the helicity independent GPDs,
giving the so called default fit of \cite{gpdff1},
and no argument is given to support this procedure
in the helicity dependent case.
Anyway, we show in Fig. 7 the comparison
of our results for the valence distribution
$\tilde H_u^v(x,0,\Delta^2)$,
evaluated at $\Delta^2 = -0.1$ GeV$^2$,
evolved to NLO from the hadronic scale up to
$Q^2=4$ GeV$^2$, with the ansatz
proposed in \cite{gpdff1}, evaluated in the same
kinematical conditions.
It is seen that the latter and the present model
predict a similar scenario. The very low $x$
behavior is governed in both cases by the Regge
phenomenology, which is left here
free to vary between the two full lines of the figure,
while it is fixed in Ref. \cite{gpdff1} in the way
described above, producing a big enhancement
at low values of $-\Delta^2$.

Summarizing this section,
we have developed here a scheme which provides us
with the helicity-dependent GPD $\tilde H$ in the full $x$ range.
This is obtained thanks to the constituent quark
structure, implemented dressing the three quarks of a CQM,
where initially only the DGLAP region of GPDs was accessible.
This is an important development, a prerequisite for
any attempt to calculate cross sections and asymmetries
of related processes.
The next step of our studies will be indeed to use the obtained
GPDs for the evaluation of cross sections.

\section{Conclusions}

A thorough analysis of both polarized
and unpolarized data have shown,
in previous work, that constituent quarks cannot be considered
elementary when studied with high-energy probes.
Certainly evolution is needed, i.e. the constituent quarks at the
hadronic scale have to be undressed by incorporating bremsstrahlung
in order to reach the Bjorken regime. Besides, the constituent quarks
should be endowed with soft structure in order to approach the data.
Thus, the constituent quarks, when under scrutiny by high-energy
probes,
seem to behave
as complex systems, with a very different
behavior from the current quarks of the basic theory.
These features, which we found in structure functions, have also
been recently discussed in form factors \cite{psr}.

The formalism of composite
constituent quarks has been recently applied
to the study of generalized parton distributions,
developing a formalism which expresses the hadronic GPDs
in terms of constituent quarks GPDs by means of appropriate
convolutions \cite{prd}.
The approach has been extended here to treat helicity-dependent
GPDs. To our knowledge, this represents the
first calculation which gives, in a constituent
quark framework, estimates of
helicity-dependent GPDs in the ERBL region.
Following the spirit of Ref. \cite{prd}, this is done
incorporating phenomenological features
of various kinematical regimes.
In particular, use has been made of the Radyushkin's
factorization ansatz, thus our constituent quark GPDs are defined in terms
of the product of three functions: i) the constituent quark
polarized structure
function, where we use the parameterization of \cite{scopetta2}
following the ACMP proposal \cite{acmp}; ii)
Radyushkin's double distributions \cite{radd}; iii)
constituent quark form factor as suggested in Ref. \cite{psr}.
Once the helicity-dependent
GPDs are defined in this way, we have developed the scheme
to incorporate them into any nucleon model
by appropriate convolution. In order to show the type of predictions
to which
our proposal leads, we have used here, as an illustration,
the IK \cite{ik} model. However, in this latest step
of our scheme, any
non-relativistic (or relativized) model can be used to define the hadronic
GPDs.

As in the unpolarized case \cite{prd},
the present scheme transforms
a hadronic model, in whose original description only
valence quarks appear, into one containing
all kinds of partons (i.e., quarks, antiquarks and gluons).
Moreover, the starting model produces no structure in
the ERBL region, while after the structure of the constituent quark
has been incorporated, it does. The completeness
of the $x$-range,
for the allowed
$\Delta^2$ and $\xi$
( $\Delta^2 << m^2$ and
$\xi^2 \ll 1$ ), of the present description,
is a prerequisite for
the calculation of cross-sections and other observables in a wide
kinematical range.
Relativistic corrections,
which permit to access a wider kinematical region, could be included
in the approach.

We have reminded the reader
of the calculation for the forward polarized
structure functions to see how in this case,
where experimental data are available,
our scheme leads, even with a naive quark model,
to a reasonable description of the data.
Thereafter, we have proceeded to calculate the GPDs of physical interest
to guide the preparation and analysis of future experiments.
Our calculation seems to be consistent with an ansatz recently
suggested in Ref. \cite{gpdff1}.

This work is the continuation of an attempt to
describe the properties of hadrons
in different kinematical and dynamical scenarios.
Although our description can
never be a substitute of Quantum Chromodynamics, it may serve
to guide experimenters to
physical processes where the theory might show interesting features, worthy
of a more fundamental effort.

\acknowledgements

Useful comments by B. Pire are gratefully acknowledged.
S.S. thanks the Department of Theoretical Physics of the Valencia
University, where part of this work has been done, for a warm hospitality
and financial support.
This work is supported in part by GV-GRUPOS03/094,
MCYT-FIS2004-05616-C02-01 and by
the Italian MIUR
through the PRIN Theoretical Studies of the Nucleus
and the Many Body Systems.

\newpage
\appendixonfalse
\section*{Figure Captions}

\vspace{1em}\noindent
{\bf Fig. 1}:
The handbag contribution to the DVCS process
in the present approach.


\vspace{1em}\noindent
{\bf Fig. 2}:
The GPD $\tilde H$ for the flavor $u$,
for $\Delta^2=-0.1$ GeV$^2$ and $\xi=0.1$,
at the momentum scale of the
model.
Dashed curve: result in the pure
Isgur and Karl model, Eq.
(\ref{iku}).
The small tails
which are found in the forbidden regions,
$x \le \xi$ and  $x \ge 1$,
represent the amount
of support violation of the approach.
Full curve: the complete result of the present
approach, Eq. (\ref{main}).
The two curves are obtained in correspondence
of the two extreme values of the Regge intercept
\cite{Heimann-Ellis},
$a_f=0$ (upper curve at low $x$) and $a_f=0.5$.

\vspace{1em}\noindent
{\bf Fig. 3}:
For the $\xi$ values
which are allowed
at $\Delta^2 = -0.1$ GeV$^2$,
$\tilde H_u(x,\xi,\Delta^2)$,
evaluated using our main equation, Eq. (\ref{main}),
is shown for
four different values of $x$,
at the momentum scale of the model.
From top to bottom,
the dash-dotted line represents the GPD at $x=0.05$,
the full line at $x=0.1$, the dashed line at $x=0.2$,
and the long-dashed line at $x=0.4$.
The curves are obtained in correspondence
of the value of the Regge intercept
$a_f=0.5$ (see text).

\vspace{1em}\noindent
{\bf Fig. 4}:
The $x$ and $\xi$ dependences of
$\tilde H_u(x,\xi,\Delta^2)$, for $\Delta^2 = -0.1$ GeV$^2$,
at the momentum scale of the model.
The value of the Regge intercept has been taken
to be $a_f=0.5$ (see text).

\vspace{1em}\noindent
{\bf Fig. 5}:
We show the structure function $xg_1^p(x,Q^2)$ obtained
at $Q^2 = 10$ GeV$^2$ by evolving at NLO the model calculation \cite{ik}
considering the structure of the constituents,
for the two extreme Regge behaviors
mentioned in the text
($a_f =0$ is the upper curve, here and also in the following figure).
The data from refs. \cite{emc}
(full dots) and \cite{smc} at $Q^2 \approx 10$ GeV$^2$ are also shown.

\vspace{1em}\noindent
{\bf Fig. 6}:
The structure function  $xg_1^n(x,Q^2)$ for the neutron
evolved at NLO
to
$Q^2= 5$ GeV$^2$, for the two extreme Regge behaviors
mentioned in the text, is shown by the two full curves.
The data from refs.
\cite{e154} at $Q^2= 5$ GeV$^2$ are also shown.

\vspace{1em}\noindent
{\bf Fig. 7}:
Comparison of the results
of the present approach for
the valence distribution
$\tilde H_u^v(x,0,\Delta^2)$ (full curves),
evaluated at $\Delta^2 = -0.1$ GeV$^2$,
evolved to NLO from the hadronic scale up to
$Q^2=4$ GeV$^2$, with the ansatz
proposed in \cite{gpdff1} (dotted curve), evaluated in the same
kinematical conditions.
The two full curves are obtained in correspondence
of the two extreme values of the Regge intercept
\cite{Heimann-Ellis},
$a_f=0$ (upper curve at low $x$) and $a_f=0.5$.


\newpage

\begin{figure}[ht]
\includegraphics{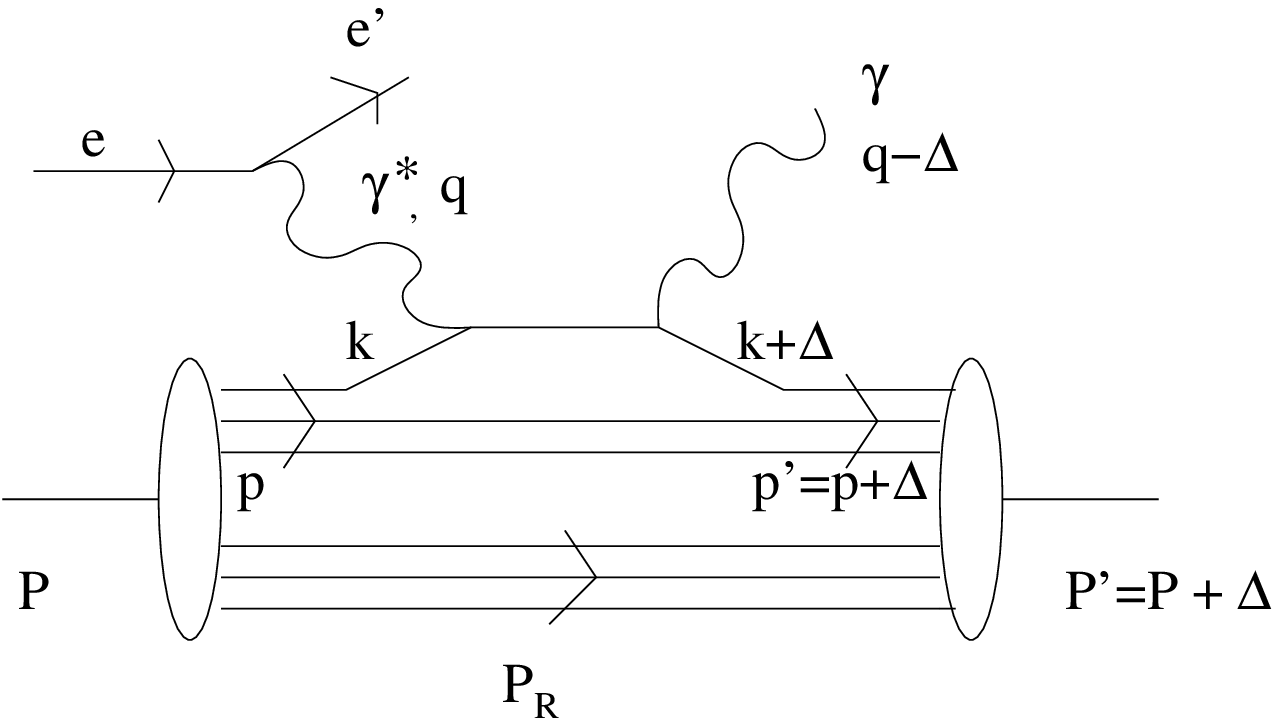}
\vspace{3cm}
\caption{}
\end{figure}




\newpage

\begin{figure}[h]
\vspace{20.0cm}
\includegraphics{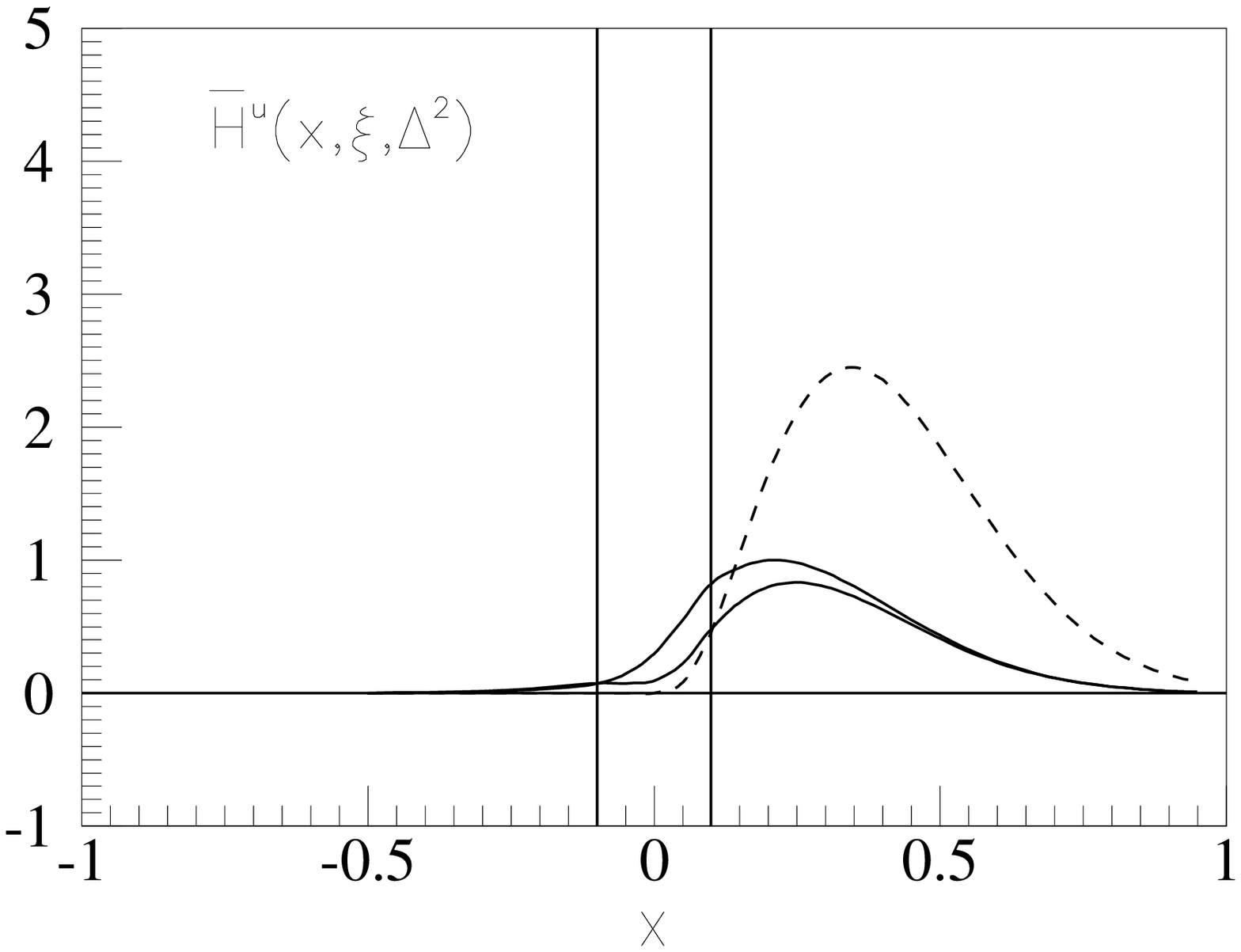}
\caption{}

\end{figure}

\newpage

\begin{figure}[h]
\vspace{24.cm}
\includegraphics{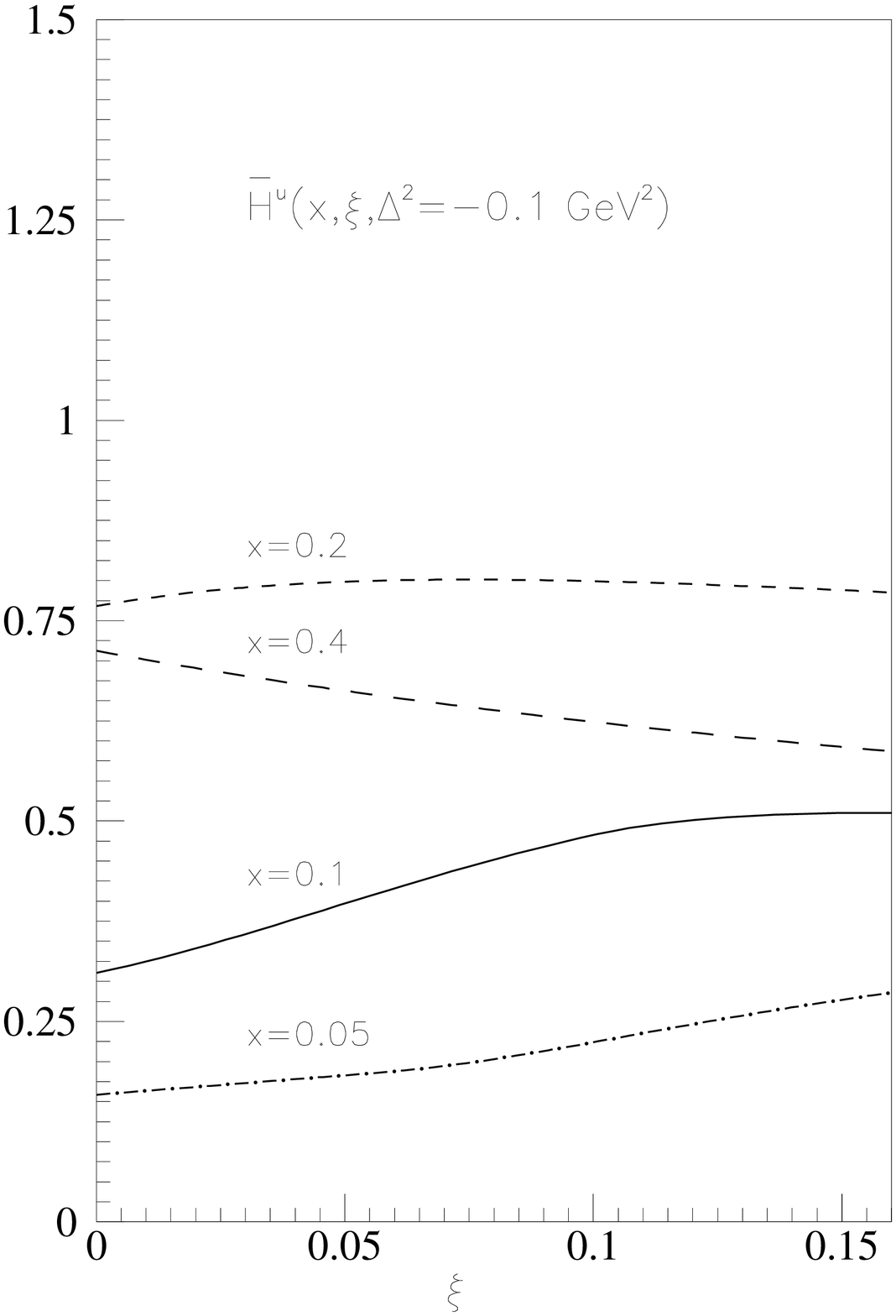}
\caption{}

\end{figure}

\newpage

\begin{figure}[h]
\vspace{12.cm}
\includegraphics{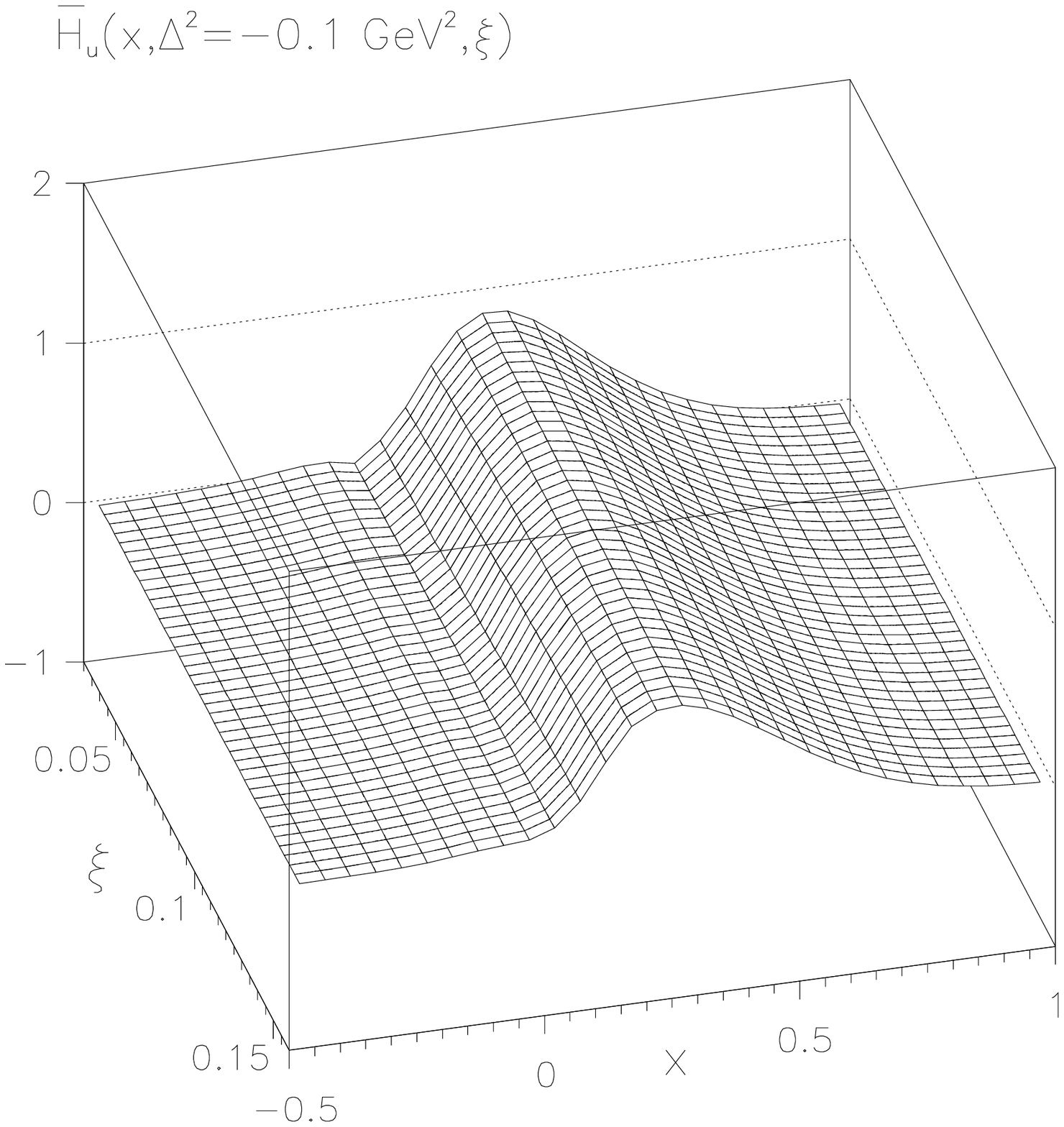}
\caption{}
\end{figure}

\newpage

\begin{figure}[h]
\vspace{12.cm}
\includegraphics{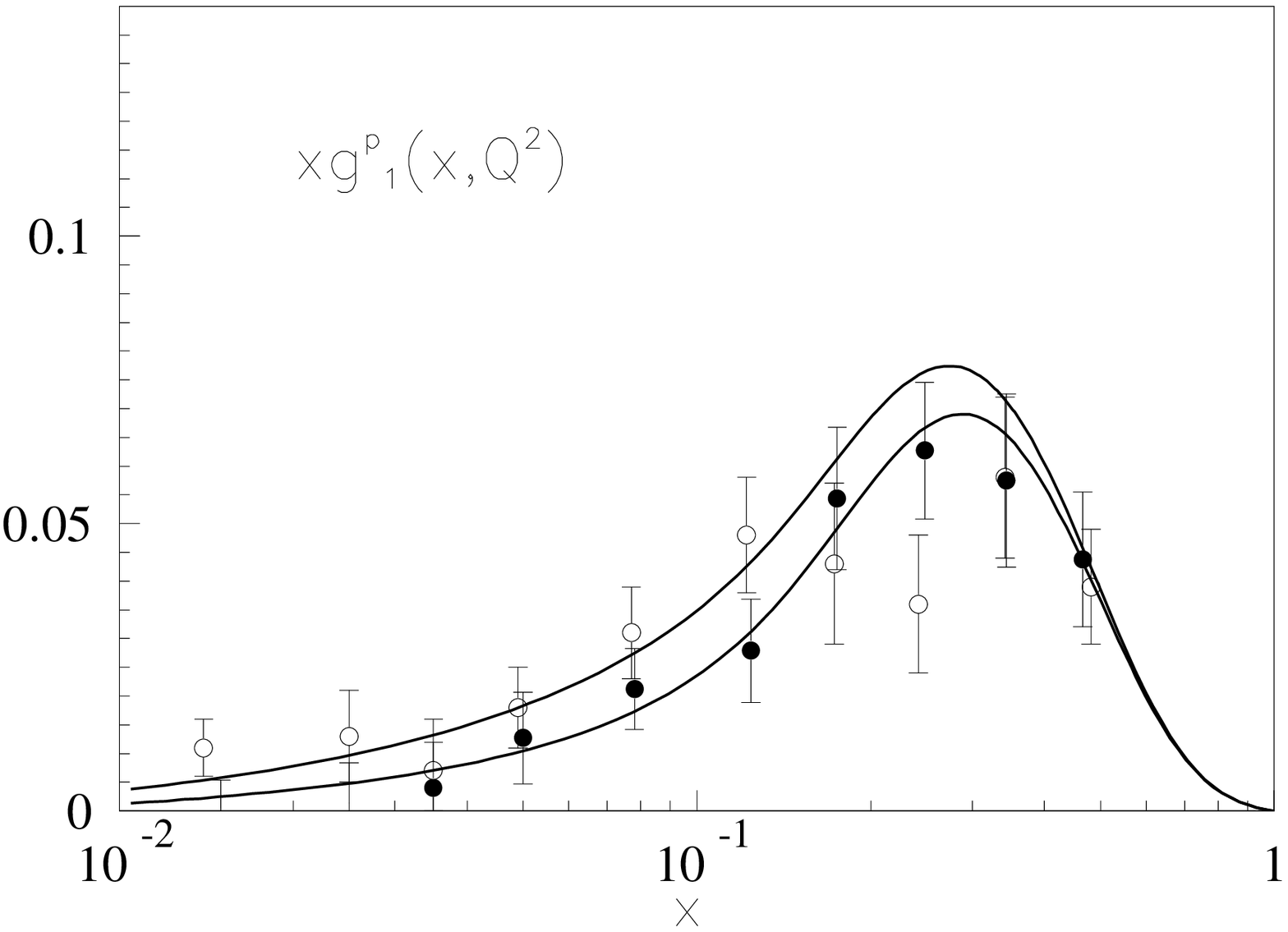}
\caption{}
\end{figure}

\newpage

\begin{figure}[h]
\vspace{8.2cm}
\includegraphics{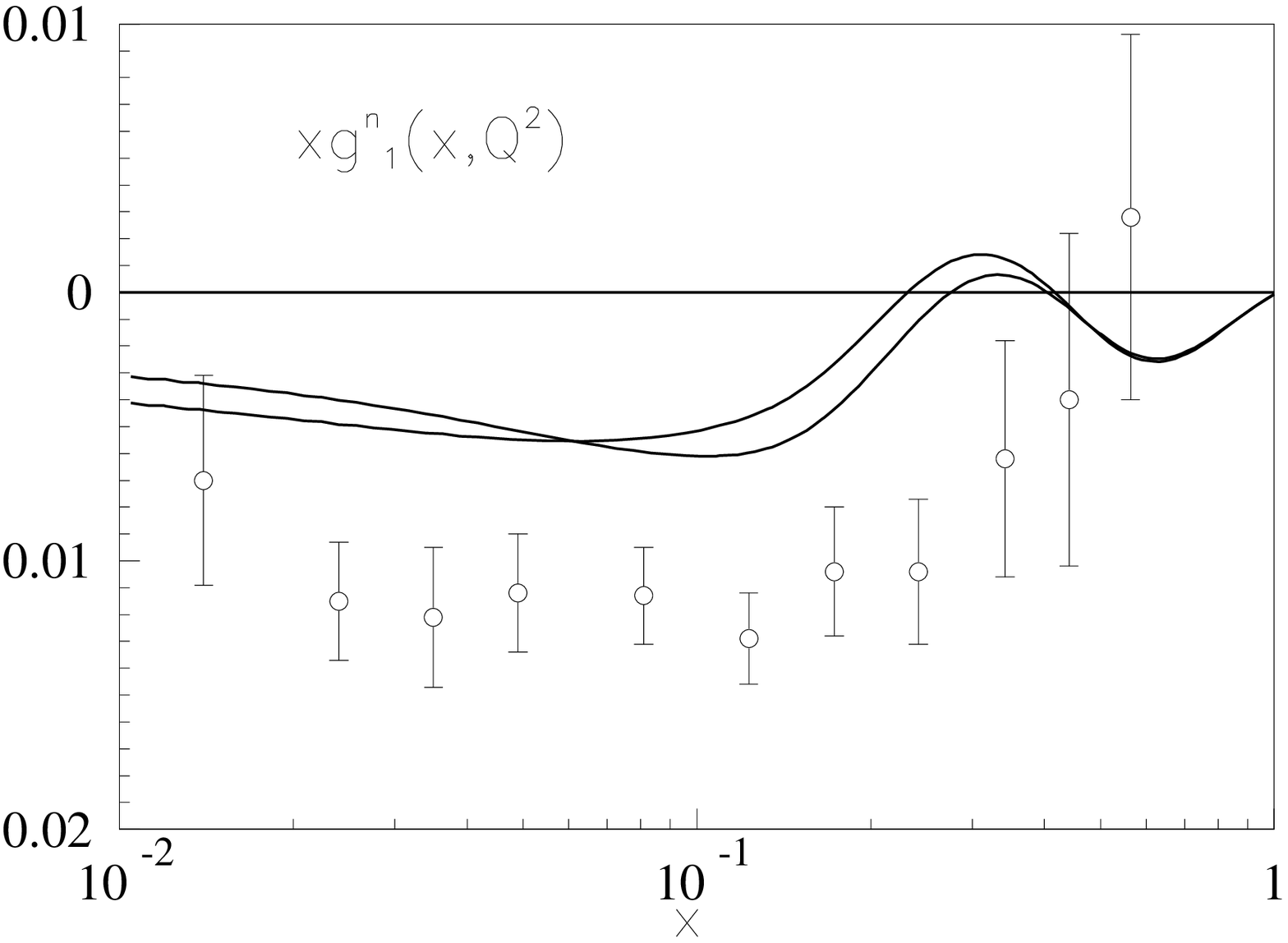}
\caption{}
\end{figure}

\newpage

\begin{figure}[h]
\vspace{8.2cm}
\includegraphics{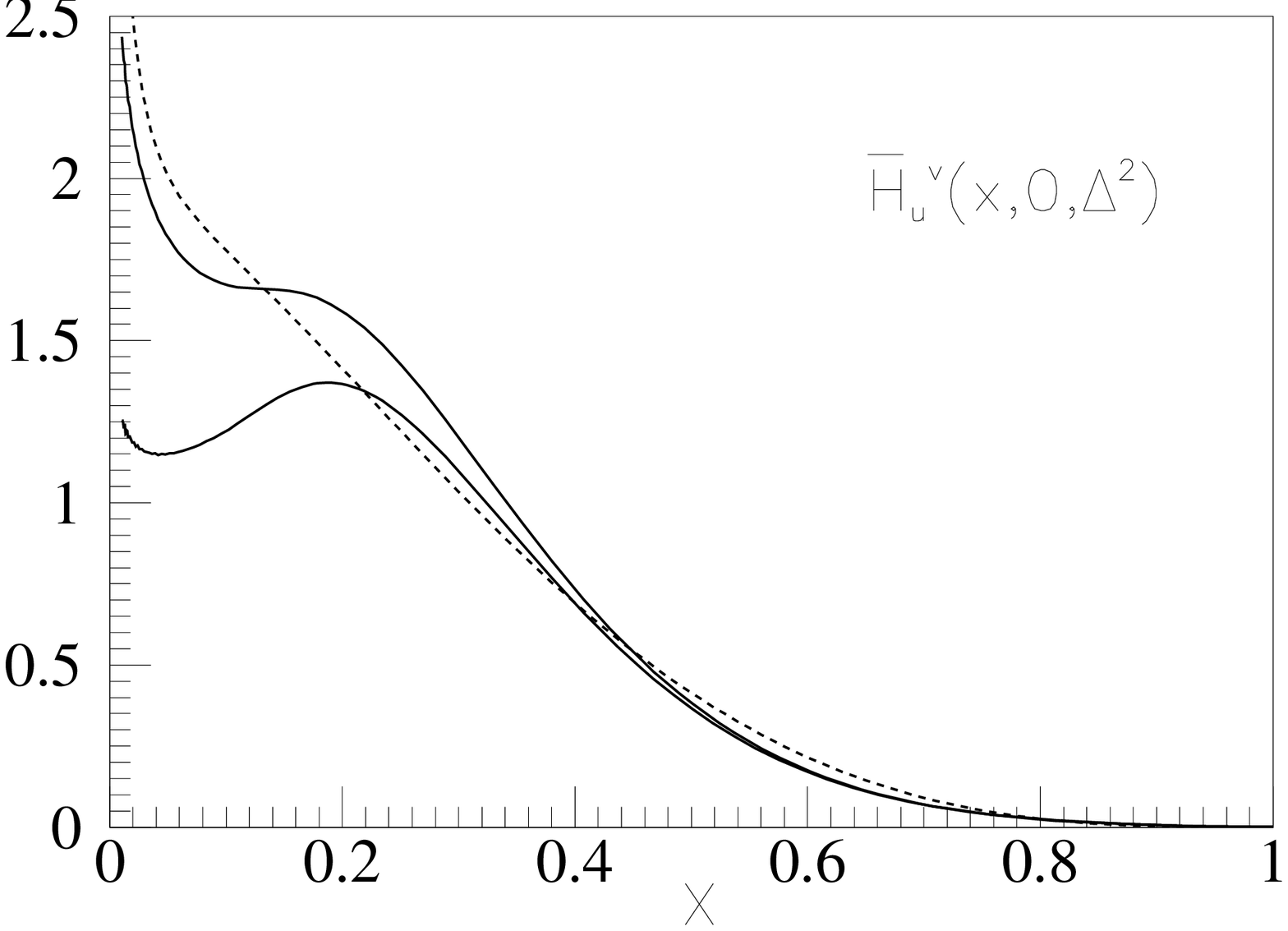}
\caption{}
\end{figure}


\begin{thebibliography}{99}
\bibitem{first} D. M\"uller, D. Robaschik, B. Geyer, F.M. Dittes,
and J. Ho\v{r}ej\v{s}i, Fortsch. Phys. 42, 101 (1994); hep-ph/9812448.
\bibitem{due}
A. Radyushkin, Phys. Lett. B 385, 333 (1996);
Phys. Rev. D 56, 5524 (1997).
\bibitem{tre} X. Ji, Phys. Rev. Lett. 78, 610 (1997);
Phys. Rev. D 55, 7114 (1997).
\bibitem{md} M. Diehl, Phys. Rept. 388, 41 (2003).
\bibitem{jaffe} R.L. Jaffe and A.V. Manohar, Nucl. Phys. B 337, 509 (1990).
\bibitem{burk} M. Burkardt, Phys. Rev. D 62, 071503 (2000);
Int. J. Mod. Phys. A 18, 173 (2003);
M. Diehl, Eur. Phys. J. C 25, 223 (2002);
J.P. Ralston and B. Pire,  Phys. Rev. D 66, 111501 (2002);
A.V. Belitsky and D. M\"uller, Nucl. Phys. A711, 118 (2002);
X. Ji, Phys. Rev. Lett. 91, 062001 (2003).
\bibitem{ceb} B.A. Mecking, Nucl. Phys. A711, 330c (2002);
K. Rith, Nucl. Phys. A711, 336c (2002);
D. von Harrach, Nucl. Phys. A711, 342c (2002).
\bibitem{dvcs} P.A. Guichon and M. Vanderhaeghen,
Prog. Part. Nuc. Phys. 41, 125 (1998).
\bibitem{meln} X. Ji, W. Melnitchouk, and X. Song,
Phys. Rev. D 56, 5511 (1997);
I.V. Anikin, D. Binosi, R. Medrano, S. Noguera,
and V. Vento, Eur. Phys. J. A 14, 95 (2002).
\bibitem{pog} K. Goeke, M.V. Polyakov, and M. Vanderhaeghen,
Prog. Part. Nucl. Phys. 47, 401 (2001).
\bibitem{goe1} V.Yu. Petrov, P.V. Pobylitsa, M.V. Polyakov,
I. Bornig, K. Goeke, and C. Weiss, Phys. Rev. D 57, 4325 (1998);
M. Penttinen, M.V. Polyakov, and K. Goeke, Phys. Rev. D 62, 014024 (2000).
\bibitem{mill} B.C. Tiburzi and G.A. Miller,
Phys. Rev. C 64, 065204 (2001);
Phys. Rev. D 65, 074009 (2002).
\bibitem{muk} A. Mukherjee and M. Vanderhaeghen,
Phys. Rev. D 67, 085020 (2003).
\bibitem{lukas} L. Theussl, S. Noguera, and V. Vento,
Eur.Phys.J.A20, 483 (2004).
\bibitem{freund}
A. Freund and V. Guzey, Phys. Lett. B 462, 178 (1999);
L. Frankfurt, A. Freund, V. Guzey, and M. Strikman,
Phys. Lett. B 418, 345c (1998).
\bibitem{rad1} A.V. Radyushkin,
Phys. Lett. B 449, 81 (1999);
I.V. Musatov and A.V. Radyushkin,
Phys. Rev. D 61, 074027 (2000).
\bibitem{scha2}
A.V. Belitsky, D. M\"uller, L. Niedermeier, and A. Sch\"afer
Phys. Lett. B 474, 163 (2000), and references therein;
A. Freund and M.F. McDermott, Eur. Phys. J. C 23, 651 (2002),
and references therein.
\bibitem{scha1}
A.V. Belitsky and D. M\"uller,
Nucl. Phys. B589, 611 (2000);
N. Kivel, Maxim V. Polyakov, A. Sch\"afer, and O.V. Teryaev,
Phys. Lett. B 497, 73 (2001);
A.V. Belitsky, A. Kirchner, D. M\"uller, and A. Sch\"afer,
Phys. Lett. B 510, 117 (2001).
\bibitem{pire} D.Yu. Ivanov, B. Pire, L. Szymanowski, and O. Teryaev,
Phys. Lett. B 550, 65 (2002).
\bibitem{pape} G. Parisi and R. Petronzio, Phys. Lett. B 62, 331 (1976);
R.L. Jaffe and G.G. Ross, Phys. Lett. B 93, 313 (1980).
\bibitem{grv} M. Gl\"uck, E. Reya, and A. Vogt, Eur. Phys. J. C 5, 461
(1998) and references therein.
\bibitem{trvv} M. Traini, V. Vento, A. Mair, and
A. Zambarda, Nucl. Phys. A614, 472 (1997).
\bibitem{epj} S. Scopetta and V. Vento,
Eur. Phys. J. A 16, 527 (2003).
\bibitem{ik} N. Isgur and G. Karl, Phys. Rev. D 18, 4187 (1978);
Phys. Rev. D 19, 2653 (1979).
\bibitem{bpt} S. Boffi, B. Pasquini, and M. Traini,
Nucl. Phys. B649, 243 (2003).
\bibitem{kroll}
M. Diehl, T. Feldmann, R. Jakob, and P. Kroll,
Nucl. Phys. B596, 33 (2001);
Eur. Phys. J. C 8, 409 (1999).
\bibitem{pietro}
P. Faccioli, M. Traini, and V. Vento, Nucl. Phys. A656, 400 (1999).
\bibitem{bptv} S. Boffi, B. Pasquini, and M. Traini,
Nucl. Phys. B680, 147 (2004).
\bibitem{prd} S. Scopetta and V. Vento, Phys. Rev. D 69,
094004 (2004).
\bibitem{scopetta1} S. Scopetta, V. Vento, and M. Traini,
Phys. Lett. B 421, 64 (1998).
\bibitem{scopetta2} S. Scopetta, V. Vento, and M. Traini,
Phys. Lett. B 442, 28 (1998).
\bibitem{acmp} G. Altarelli, N. Cabibbo, L. Maiani, and R. Petronzio,
Nucl. Phys. B69, 531 (1974); G. Altarelli, S. Petrarca, and
F. Rapuano, Phys. Lett. B 373, 200 (1996).
\bibitem{psr} R. Petronzio, S. Simula, and G. Ricco,
Phys. Rev. D 67, 094004 (2003).
\bibitem{mor} G. Morpurgo, Physics 2, 95 (1965).
\bibitem{hwa} R.C. Hwa, Phys. Rev. D 22, 759 (1980);
R.C. Hwa and C.B. Yang, Phys. Rev. C 66, 025204 (2002)
and references therein.
\bibitem{radd} A.V. Radyushkin, Phys. Rev. D 59, 014030 (1999).
\bibitem{rag} A.V. Radyushkin,
in M. Shifman, (Editor): At the Frontier of Particle Physics, Vol. 2,
(World Scientific, Singapore, 2001) pp. 1037-1099,
hep-ph/0101225;
\bibitem{jig} X. Ji, J. Phys. G 24, 1181 (1998).
\bibitem{muld} P. Mulders, Phys. Rept. 185, 83 (1990).
\bibitem{fs} L.L. Frankfurt and M.I. Strikman,
Phys. Rept. 160, 235 (1988).
\bibitem{rope} M. Ropele, M. Traini and V. Vento,
Nucl. Phys. A 584, 634 (1995).
\bibitem{mun} L. Mankiewicz, G. Piller, and T. Weigl, Eur. Phys. J. C 5,
119 (1988).
\bibitem{prc} S. Scopetta, Phys. Rev. C 70, 015205 (2004).
\bibitem{csps} C. Ciofi degli Atti, S. Scopetta, E. Pace, and G. Salm\`e,
Phys. Rev. C48, R968 (1993).
\bibitem{cps} C. Ciofi degli Atti, E. Pace, and G. Salm\`e,
Phys. Rev. C46, R1591 (1992).
\bibitem{cano} F. Cano and B. Pire, Nucl. Phys. A 711, 133c (2002);
Eur. Phys. J. A 19, 423 (2004).
\bibitem{str} V. Guzey and M.I. Strikman, Phys. Rev. C 68, 015204 (2003).
\bibitem{Heimann-Ellis} R. L. Heimann, Nucl. Phys. B64 (1973) 429; J. Ellis
and M. Karliner, Phys. Lett. B213 (1988) 73.
\bibitem{kaur} R. Carlitz and J. Kaur, Phys. Rev. Lett. 38, 673 (1977);
J. Kaur, Nucl. Phys. B 128, 219 (1977);
A. Schaefer, Phys. Lett. B 208, 175 (1988).
\bibitem{abfr} G. Altarelli, R.D. Ball, S. Forte and G. Ridolfi,
 \np B 496, 337 (1997).
\bibitem{mmg} M.M. Giannini, Rep. Prog. Phys. 54, 453 (1991).
\bibitem{mainz} L.A. Ahrens et al., Phys. Rev. D 35, 785 (1987).
\bibitem{iac} R. Bijker, F. Iachello and A. Leviatan,  Ann. Phys. 236,
69 (1994); Phys Rev. C 54, 1935 (1996);
Phys. Rev D 55, 2862  (1997).
\bibitem{fp} W. Furmansky and R. Petronzio, Z. Phys. C 11, 293 (1982).
\bibitem{bfr} R.D. Ball, S. Forte and G. Ridolfi,
Phys. Lett. B 378, 255 (1996).
\bibitem{NLOP} R. Mertig, W. L. van
Neerven, Z.Phys. C70, 637 (1996); W. Vogelsang,
Phys. Rev D54, 2023 (1996).
\bibitem {Altarelli-Ross} G. Altarelli and G. G. Ross,
Phys. Lett. B 212, 391 (1988).
\bibitem{cr} F.E. Close and R.G. Roberts, Phys. Lett. B 316, 165
(1993).
\bibitem{pdg} Particle Data Group, J. Ashman {\sl et al.,} Phys. Rev. D
54, 1 (1996).
\bibitem{emc} EMC Collaboration, J. Ashman {\sl et al.,} Nucl. Phys.
B 328, 1 (1989) .
\bibitem{smc} SMC Collaboration, D. Adams {\sl et al.,}
Phys. Rev. D 56, 5330, (1997).
\bibitem{e154} E154 Collaboration, K. Abe {\sl et al.,} Phys. Rev. Lett.
79, 26 (1997).
\bibitem{nsv} S. Noguera, S. Scopetta and V. Vento,
Phys. Rev. D70, 094018 (2004).
\bibitem{gpdff1} M. Diehl, Th. Feldmann, R. Jakob, and P. Kroll,
hep-ph/0408173.
\bibitem{gpdff2} M. Guidal, M.V. Polyakov, A.V. Radyushkin, and
M. Vanderhaeghen, hep-ph/0410251.
\bibitem{parpol} J. Bl\"umlein and M. B\"ottcher, Nucl. Phys. B 636,
225 (2002).
\end{thebibliography}
\end{document}